\documentclass[preprint2]{aastex62}

\usepackage{lineno}
\usepackage{amsmath}
\graphicspath{{./}{figures/}}

\accepted{to ApJ}
\shorttitle{Clouds and Hazes in TRAPPIST-1}
\shortauthors{Fauchez et al.}
\begin{document}

\title{Impact of Clouds and Hazes on the Simulated JWST Transmission Spectra \\ of Habitable Zone Planets in the TRAPPIST-1 System}

\correspondingauthor{Thomas Fauchez}
\email{Thomas.j.fauchez@nasa.gov}

\author[0000-0002-5967-9631]{Thomas J. Fauchez}
\affiliation{NASA Goddard Space Flight Center, Greenbelt, MD, USA}
\affiliation{Goddard Earth Sciences Technology and Research (GESTAR), Universities Space Research Association, Columbia, MD, USA}
\affiliation{GSFC Sellers Exoplanet Environments Collaboration}

\author{Martin Turbet}
\affiliation{Laboratoire de M\'eteorologie Dynamique, IPSL, Sorbonne Universit\'es, UPMC Univ Paris 06, CNRS, 4 Place Jussieu, 75005 Paris, France}
\affiliation{Observatoire  Astronomique  de  l’Universit\'e  de  Gen\`eve,  Universit\'e  de  Gen\`eve,  Chemin  des Maillettes 51, 1290 Versoix, Switzerland.}

\author{Geronimo L. Villanueva}
\affiliation{NASA Goddard Space Flight Center, Greenbelt, MD, USA}
\affiliation{GSFC Sellers Exoplanet Environments Collaboration}

\author{Eric T. Wolf}
\affiliation{Laboratory for Atmospheric and Space Physics, Department of Atmospheric and Oceanic Sciences, University of Colorado Boulder, Boulder, CO, USA}
\affiliation{NASA Astrobiology Institute’s Virtual Planetary Laboratory, Seattle, WA, USA}

\author{Giada Arney}
\affiliation{NASA Goddard Space Flight Center, Greenbelt, Maryland, USA}
\affiliation{GSFC Sellers Exoplanet Environments Collaboration}

\author{Ravi K. Kopparapu}
\affiliation{NASA Goddard Space Flight Center, Greenbelt, Maryland, USA}
\affiliation{GSFC Sellers Exoplanet Environments Collaboration}

\author{Andrew Lincowski}
\affiliation{Department of Astronomy and Astrobiology Program, University of Washington, Box 351580, Seattle, WA 98195, USA}
\affiliation{NASA Astrobiology Institute’s Virtual Planetary Laboratory, Box 351580, University of Washington, Seattle, WA 98195, USA}

\author{Avi Mandell}
\affiliation{NASA Goddard Space Flight Center, Greenbelt, MD 20771, USA}
\affiliation{GSFC Sellers Exoplanet Environments Collaboration}

\author{Julien de Wit}
\affiliation{Department of Earth, Atmospheric and Planetary Sciences, Massachusetts Institute of Technology, Cambridge, MA, USA.}

\author{Daria Pidhorodetska}
\affiliation{NASA Goddard Space Flight Center, Greenbelt, Maryland, USA}
\affiliation{University of Maryland Baltimore County/CRESST II 1000 Hilltop Cir. Baltimore, MD 21250, USA}

\author{Shawn D. Domagal-Goldman}
\affiliation{NASA Goddard Space Flight Center, Greenbelt, Maryland, USA}
\affiliation{GSFC Sellers Exoplanet Environments Collaboration}

\author{Kevin B. Stevenson}
\affiliation{Space Telescope Science Institute, Baltimore, MD 21218, USA}
\affiliation{John's Hopkins APL,1100 Johns Hopkins Rd, Laurel, MD 20723}



\begin{abstract}

The TRAPPIST-1 system, \added{consisting of an ultra-cool host star having seven known} Earth-size planets will be a prime target for atmospheric characterization with JWST. 
However, the detectability of atmospheric molecular species may be  severely impacted by the presence of clouds and/or hazes.
In this work, we perform 3-D General Circulation Model (GCM) simulations with the LMD Generic model supplemented by 1-D photochemistry simulations at the terminator with the Atmos model to simulate several possible atmospheres for TRAPPIST-1e, 1f and 1g:  1) modern Earth, 2) Archean Earth, and 3) CO$_2$-rich atmospheres. JWST synthetic transit spectra were computed using the GSFC Planetary Spectrum Generator (PSG).
We find that TRAPPIST-1e, 1f and 1g atmospheres, with clouds and/or hazes, could be detected using JWST's NIRSpec prism from the CO$_2$ absorption line at $4.3\ \mu m$ in less than 15 transits  at  $3\ \sigma$ or less than 35 transits at  $5\ \sigma$.  However, our analysis suggests that other gases would require hundreds (or thousands) of transits to be detectable.
We also find that H$_2$O, mostly confined in the lower atmosphere, is very challenging to detect for these planets or similar systems if the planets' atmospheres are not in a moist greenhouse state. This result demonstrates that the use of GCMs,  self-consistently taking into account the effect of clouds and sub-saturation, is crucial to evaluate the detectability of atmospheric molecules of interest as well as \added{for} interpreting future detections in a more global (and thus robust and relevant) approach. \\
 
\end{abstract}

%


\section{Introduction} \label{sec:intro}
During the last decade, a \added{rapidly} increasing number of Earth-size planets in the so-called Habitable Zone (HZ) have been discovered. Among the most famous of them are Kepler-186f \citep{Quintana2014}, Proxima Centauri b \citep{Anglada2016}, GJ 1132b \citep{Berta-Thompson2015}, Ross 128b \citep{Bonfils2018} and the TRAPPIST-1 system \citep{Gillon2016,Gillon2017}.  The HZ is defined as the region around a star where a planet with appropriate atmospheric pressure, temperature and composition can maintain liquid water on its surface \citep{Kasting1993,Selsis2007,Kopparapu2013,Yang2014,Kopparapu2017}, which is crucial for life as we know it. However, the abundant presence of liquid water at the surface of a planet is not the only criteria that deems it to be  habitable. The planet's geophysics and geodynamics as well as its interaction with its host stars' plasma and radiation environment are also crucial parameters to determine its habitability \citep{Lammer2009}. Low-mass stars (late K and all M-dwarf stars) provide the best opportunity for detecting and characterizing habitable terrestrial planets in the coming decade. The small size of these stars allows for a greater chance of detection of terrestrial-sized planets, and planets in their compact HZ which orbit more frequently lead to a better signal-to-noise (\added{S/N}) level than planets orbiting in the HZ of Sun-like stars.  For late M-dwarfs such as TRAPPIST-1 (M8V), the \added{S/N} can be amplified by a factor up to 3 compared to stars with type earlier than M1 \citep{deWit2013}. Among the most promising systems with planets in the HZ of low mass stars is the nearby TRAPPIST-1 system, located 12~pc away, discovered by \cite{Gillon2016,Gillon2017,Luger2017}  and composed of at least seven rocky planets with three of them in the HZ. The system's host star, TRAPPIST-1, is an active late M--dwarf \citep{OMalley2017,Wheatley2017,Vida2018} whose stellar flares could bathe the planetary environment with high energy radiation and plasma, creating severe obstacles to retaining an atmosphere or sustaining habitable conditions on their surface. Despite these difficult conditions,  \cite{Bolmont2017}, \cite{Bourrier2017} and \cite{Dong2018} have argued that the TRAPPIST-1 planets could retain surface liquid water if they were formed with abundant initial water endowment. Transit-timing variation (TTVs) measurements of the TRAPPIST-1 planets by \cite{Grimm2018} have also suggested a volatile-rich composition and thus a potentially large amount of water. \\

The proximity of the TRAPPIST-1 system and the high frequency of its planetary transits \added{makes it a} prime target for temperate rocky exoplanet atmospheric characterization. The first atmospheric characterization with HST by \cite{deWit2016,deWit2018} revealed that the TRAPPIST-1 planets do not contain a cloud/haze free H$_2$--dominated atmosphere but may instead be composed of a wide variety of atmospheres dominated by  N$_2$, O$_2$, H$_2$O, CO$_2$, or CH$_4$. Following these studies,  \cite{Moran2018} have used lab measurements and a 1-D atmospheric model to show that H$_2$--dominated atmospheres with cloud/haze would better fit the spectra than clear sky H$_2$--dominated atmospheres (except for TRAPPIST-1g). However, the noise on the HST transmission spectra is on the order of hundreds of parts per million (ppm) \citep{deWit2018}. The sensitivity, spectral resolution and the wide wavelength coverage of the future James Webb Space Telescope (JWST) will be needed to address whether or not these planets have an atmosphere and to uncover clues related to their composition \citep{Barstow2016,Morley2017}. \\

Before JWST observes these planets, it is important to understand the possible composition of their atmospheres (if any) and their climate conditions. The 3-D General Circulation Models (GCMs) are the most sophisticated tools to address these questions because they can simulate tidally-locked planets, and they allow for a self-consistent and coupled treatment of all physical processes occurring in a planetary atmosphere. This is particularly important for water in its various thermodynamic phases which is responsible for the water-vapor greenhouse feedbacks  and the sea-ice albedo. Both are strong and positive effects amplifying temperature perturbations to the climate system in either direction. The interaction between water vapor and the 3-D atmospheric dynamics controls the relative humidity of the atmosphere, and ultimately the strength of a planet's water vapor greenhouse effect \citep{Pierrehumbert1995}. Similarly, the spatial distributions of sea-ice, snow coverage, and clouds largely determines the planetary albedo.\\

One-dimensional (1-D, vertical) models struggle to simulate rocky planets within the habitable zones of low-mass stars (late-K and all M-dwarfs) in synchronous rotation \citep{Leconte2015,Barnes2017}, \added{yet} it is precisely these planets, such as the TRAPPIST-1 system, that have deep transits and shorter orbits that allow for near-term atmospheric characterization \citep{Kaltenegger2009,Snellen2013} with JWST or future ground-based observatories such as E-ELT, GMT or TMT.  In this situation, one side of the planet is permanently exposed to starlight, while the other side is condemned to permanent darkness. Tidally-locked planets in the HZ of low-mass stars usually have rotation periods that are much longer than the one of Earth, leading to a weaker Coriolis force. Instead of having primarily zonal flows with mid-latitude jets like on Earth, \added{modeling studies suggest that these planets have}  sub-stellar to anti-stellar radial flow aloft with strong rising motions on the permanent day-side, and subsiding motions on the permanent night-side \citep{Joshi2003, Merlis2010}. Only 3-D climate models can capture these motions that have strong effects on the climate system.  \cite{Yang2013} showed that slow and synchronously rotating planets have thick clouds near the substellar point, drastically increasing the planetary albedo, and inhibiting \added{the planet from entering}  a runaway greenhouse \added{state even at much} higher incident stellar fluxes compared to an Earth-Sun twin. 3-D models have also been used to study the spatial variability of chemical species of rocky exoplanets and have found that significant chemical gradients exist between the day and night sides of slow rotating planets \citep{Chen2018}. Finally, several studies  \citep{Hu2014,Way2017,DelGenio2019} have shown the importance of accounting for the ocean heat transport for slow rotating habitable planets. Each of these processes bears a great impact on a planet's climate and can only be adequately portrayed through the use of 3-D models. \added{However, 3-D models require a larger set of initial and boundary conditions, which are not well known for exoplanets, while simpler but faster 1-D  models can explore a larger parameter space; therefore the two approaches are complementary.}\\

Some 3-D GCM simulations of the TRAPPIST-1 planets have already been performed. \cite{Wolf2017} and \cite{Turbet2018} have shown that TRAPPIST-1e is the most likely planet to be habitable, based on the result that it can retain liquid water on its surface for a large set of atmospheric compositions and thicknesses. Moreover, \added{results suggest} that a few bars of \added{surface} CO$_2$ are needed to maintain ice-free surfaces on TRAPPIST-1f and g. \cite{Grimm2018} has found that while TRAPPIST-1e may have a large rocky interior, TRAPPIST-1f and -1g are likely to be volatile rich. Note that 1-D climate model simulations have also been used for TRAPPIST-1 planets \citep{Morley2017, Lincowski2018,Lustig_Yaeger2019} with the limitations of this approach described earlier. For instance, in their simulated transmission spectra \cite{Morley2017}, considers clear sky atmospheres, while it is not realistic when H$_2$O or CO$_2$ are in the atmosphere and could eventually form clouds, or CH$_4$ and H$_2$SO$_4$ which could form organic and sulfuric hazes. In \cite{Lincowski2018}, water cloud optical thicknesses have not been used consistently with the water vapor mixing ratio. \cite{Lustig_Yaeger2019} results suggest that NIRSpec is the most favorable JWST instrument to characterize TRAPPIST-1 planet's atmosphere and that only few transits would be needed to detect CO$_2$.\\

Clouds or other aerosols such as photochemical hazes could have a large impact on both the climate and the detectability of spectral features through transmission spectroscopy. Atmospheric conditions favoring the presence of clouds and/or hazes could severely impact the observed transmission spectra by flattening spectral lines.  This phenomenon has been shown to be widespread in observations of larger planets with clouds, such as super-Earth GJ~1214b \citep{Kreidberg2014}, gaseous giant WASP-12b \citep{Wakeford2017}, and WASP-31b \citep{Sing2016}.  It has also been observed for hazes, for example on WASP-6b \citep{Nikolov2015}  and HAT-P-12b \citep{Sing2016}.  Furthermore, \cite{Arney2017} simulated JWST observations for a hazy Archean Earth orbiting around the M4 dwarf star GJ~876 using the \cite{Deming2009} JWST simulator. The spectra were computed with an atmospheric model coupled to the Spectral Mapping Atmosphere Radiative Transfer model (SMART, \cite{Meadows1996}). They showed that the hazes flatten the spectrum and reduce the relative spectral impact of gaseous absorption in the JWST NIRISS bandpass. Hazes can significantly impact JWST spectra, and accounting for them can improve observational strategies of potentially haze-rich worlds.\\

In this paper, we use a 3-D GCM adapted for the TRAPPIST-1 planets \citep{Turbet2018} to explore how aerosol formation, including H$_2$O liquid and ice clouds, CO$_2$ clouds and photochemical organic hazes, impact the atmospheres and the simulated transmission spectra of TRAPPIST-1e, 1f and 1g, with a focus here on the scenarios where these planets are habitable.
The TRAPPIST-1 system is a natural laboratory for studying haze and cloud formation because the planets receive a wide range of incident stellar fluxes. Therefore, the results of our study can be applied to a wide range of Earth-sized planets orbiting M-dwarfs. Hazes and clouds are notoriously difficult to model in 1-D, which motivated us to use a GCM in this work. However, the chemistry that impacts cloud and haze formation is difficult to simulate in 3-D, mainly because it requires large computing time. To date, the best solution is a nested set of models that leverage the ability of 1-D models to simulate photochemistry/hazes and the ability of 3-D GCMs to simulate clouds. In this work, we have sequentially connected (i) the 3-D General Circulation Model simulations accounting for cloud formation with (ii) a 1-D radiative-convective photochemical model accounting for the formation of hazes, along with (iii) a transit transmission spectra generator to model JWST observations.

The main purpose of this paper is to explore whether and how clouds and photochemical hazes can affect our ability to characterize the atmospheric composition of habitable planets around TRAPPIST-1.We chose to explore three main types of habitable planet atmospheres, representative of habitable planets known to exist and to have existed in the Solar System:
\begin{enumerate}
    \item Modern Earth: This is the best example we have of a habitable planet. It is also the most widespread benchmark for habitable planets in the literature \citep{Barstow2016,Morley2017,Lincowski2018}.
    \item Archean Earth: This case is representative of the early Earth (during the Archean epoch), at a time when Earth had oceans of liquid water, despite a different atmosphere (i.e. both CO$_2$ and CH$_4$-rich) from today's atmosphere. For this case of a habitable planet, we used different scenarios of Archean atmospheres from \cite{Charnay2013}.
    \item Planet with a thick CO$_2$-dominated atmosphere: This case is likely representative of the early Earth (during the Hadean epoch), early Venus, and early Mars, at a time when Martian valley networks and lakes were formed \citep{Haberle2017,Kite2019}.
\end{enumerate}

The paper is structured as follows: Section \ref{sec:method} discusses the method and the tools used in this study to simulate both the climate and the transmission spectra of TRAPPIST-1 planets in the HZ. Sections~\ref{sec:Earth} to \ref{sec:CO2} successively present the climate and JWST transmission spectra for the three types of habitable planets introduced above (modern Earth, Archean Earth and planets with a thick CO$_2$-dominated atmosphere). The sections have been ordered by degree of complexity.  In Section~\ref{sec:Earth}, we focus on simulated atmospheres with boundary conditions based on the modern Earth, highlighting the effect of clouds and photochemical molecular species. In Section~\ref{sec:Archean}, we focus on a simulated atmospheres based on Archean Earth boundary conditions, highlighting the effect of clouds, photochemical molecular species, and photochemical hazes. In Section~\ref{sec:CO2}, we focus on CO$_2$ dominated atmospheres, highlighting the effect of H$_2$O and CO$_2$ clouds. Discussions of our results are provided in Section \ref{sec:discussion}, with a particular emphasis on the detectability of H$_2$O. Finally, conclusions and perspectives are presented in Section~\ref{sec:conclusions}.

\section{Method: from climate to spectra} \label{sec:method}
\subsection{Simulation of the climate}

The Laboratoire de M\'et\'eorologie Dynamique Generic (LMD-G, \cite{Wordsworth2011}) model is the (exo)planetary version of \added{the Laboratoire de M\'et\'erologie Dynamique zoom (LMDz)}, a General Circulation Model historically built using Mars \citep{Forget1999} and Earth \citep{Hourdin2006} LMD GCMs. This is a versatile GCM, able to handle a wide range of temperatures and surface pressures as well as various condensates (e.g. H$_2$O, CO$_2$, CH$_4$, N$_2$). Numerous studies have taken advantages of the GCM's versatility to model planetary atmospheres in the Solar System and beyond \citep{Wordsworth2011,Wordsworth2013,Wordsworth2015,Leconte2013a,Leconte2013b,Charnay2013,Charnay2015a,Charnay2015b,Bolmont2017,Turbet2016,Turbet2017a,Turbet2017b,Turbet2018}.

\subsubsection{Radiative transfer}
LMD-G uses a generalized radiative transfer algorithm for the absorption and scattering by the atmosphere, the clouds and the surface from the far-infrared to visible range \citep{Wordsworth2011}. The scattering effects of the atmosphere and the clouds are parameterized through a two-stream scheme \citep{Toon1989} using the method of \cite{Hansen1974}.
Absorption coefficients are computed with the correlated-k distribution method \citep{LacisOinas1991} using absorption lines from HITRAN~2008 \citep{Rothman2009}. The collision-induced and dimer absorptions \citep{Wordsworth2010a,Richard2012} and the sublorentzian profiles \citep{Perrin1989} were computed as in \cite{Charnay2013} and \cite{Wordsworth2013}. Present-day Earth, Archean Earth and CO$_2$-dominated atmospheres absorption coefficients were computed as in \cite{Leconte2013a}, \cite{Charnay2013} and \cite{Wordsworth2013}, respectively. Between 36 and 38 spectral bands are considered in the shortwave and between 32 and 38 in the longwave range. Sixteen non-regularly spaced grid points were adopted for the g-space integration, with ``g" the cumulative distribution function of the absorption data for each band. \\

TRAPPIST-1 emission spectra were computed using the synthetic BT-Settl spectrum \citep{Rajpurohit2013} assuming a temperature of 2500~K, a surface gravity of $10^3\ m\cdot s^{-2}$ and a metallicity of 0 dex, as in \cite{Turbet2018}. For planets orbiting an ultra-cool star like TRAPPIST-1, the bolometric albedo of water ice and snow is significantly lowered \citep{Joshi2012, vonParis2013,Shields2013} due to the shape of its reflectance spectrum as explained in \cite{Warren1980,Warren1984}. To account for this effect, LMD-G computes the wavelength-dependent albedo of water ice and snow following a simplified albedo spectral law, previously calibrated to match the ice and snow bolometric albedo of 0.55 around a Sun-like star \citep{Turbet2016}. Around TRAPPIST-1, the average bolometric albedo for water ice and snow has been estimated to be $\sim 0.21$ \citep{Turbet2018}. 

\subsubsection{Microphysics}
For each of the simulations performed in this study, water vapor was treated as a variable species. In other words, the relative water vapor humidity is set free and super-saturation is not permitted by the LMD-G moist convective adjustment scheme \citep{Leconte2013b}. Water phase transitions, such as melting, freezing, condensation, evaporation and sublimation as well as water precipitation, were also considered. Water precipitation was computed with the scheme of \cite{Boucher1995}. Similarly, the possible condensation and/or sublimation of CO$_2$ in the atmosphere (and on the surface) has been taken into account  but not the radiative effect of CO$_2$ ice clouds because their scattering greenhouse effects \citep{Forget1997} are low around cool dwarf stars such as TRAPPIST-1 \citep{Kitzmann2017} and are also limited by partial cloud coverage \citep{Forget2013}. When/where   H$_2$O and/or CO$_2$ condenses, evaporates, or sublimates, the effect of latent heat also is taken into account. CO$_2$ and H$_2$O cloud particle sizes were estimated from the amount of condensed material and the number density of cloud condensation nuclei (CCN). CCNs have been set up to a constant value of $10^6\ kg^{-1}$ for liquid water clouds, $10^4\ kg^{-1}$ for water ice clouds \citep{Leconte2013a} and $10^5\ kg^{-1}$ for CO$_2$ ice clouds \citep{Forget2013} everywhere in the atmosphere. Ice particles and liquid droplets were sedimented following a Stokes law from \cite{Rossow1978}. 

\subsubsection{Climate simulations of TRAPPIST-1e, 1f and 1g}
In this work, we have performed GCM simulations of the TRAPPIST-1 planets using planetary properties from \cite{Gillon2017} and \cite{Grimm2018}. A summary of the planetary properties used in this work is provided in Table \ref{tab:TRAPPIST1}. We have considered here only the three planets located in  the classical HZ \citep{Kopparapu2013}, namely TRAPPIST-1e, f and g, which are all assumed to be fully covered by a 100~m deep ocean (aqua-planets \added{with no orography}) with a thermal inertia of $12000~J\cdot m^{-2}\cdot K^{-1}\cdot s^{-2}$ with no ocean heat transport (OHT). 

For such a close-in system, the planets are believed to be in synchronous rotation \citep{Turbet2018}. In a synchronous rotation regime, thermal inertia should only affect the variability of the atmosphere.
The  horizontal resolution adopted for all the simulations is a  $64\times 48$ coordinates in longitude $\times$ latitude (e.g., $5.6^{\circ} \times 3.8^{\circ}$). In the vertical direction, the atmosphere is discretized in 26 distinct layers (model top at 10$^{-5}$ bar) using the hybrid $\sigma$ coordinates while the ocean is discretized in 18 layers.
The dynamical, physical and radiative transfer time steps have been set up to set to 90, 900 and 4500~s, respectively. 

For each of the three planets, the atmospheric configurations below have been modeled. The motivation for their selection is to highlight the impact of the following aerosols: H$_2$O (liquid and ice), CO$_2$ ice, and photochemical organic hazes.
\begin{itemize}
\item Modern Earth-like (1~bar of N$_2$, 376~ppm of CO$_2$): Expected to form H$_2$O clouds.
\item Archean Earth-like:
\begin{itemize}
    \item \cite{Charnay2013} case A (0.998 bar of N$_2$, 900~ppm of CO$_2$, 900~ppm of CH$_4$: Expected to form H$_2$O clouds and photochemical hazes (see subsection \ref{sub:ATMOS})
    \item \cite{Charnay2013} case B (0.988 bar of N$_2$, 10,000~ppm of CO$_2$, 2,000~ppm of CH$_4$: Expected to form H$_2$O clouds and photochemical hazes (see subsection \ref{sub:ATMOS})
    \item \cite{Charnay2013} case C (0.898 bar of N$_2$, 100,000~ppm of CO$_2$, 2,000~ppm of CH$_4$: Expected to form H$_2$O clouds
\end{itemize}
\item CO$_2$-dominated atmospheres:  
\begin{itemize}
    \item 1 bar surface pressure: Expected to form H$_2$O and CO$_2$ clouds
    \item 10 bars surface pressure: Expected to form H$_2$O and CO$_2$ clouds
\end{itemize}
\end{itemize}

Each simulation was run until the radiative equilibrium had been reached at the top of the atmosphere (TOA), typically after a couple of tens of Earth years. Simulations that lead to unstable CO$_2$ surface collapse, i.e. when the rate of CO$_2$ surface condensation reached a positive constant \citep{Turbet2017b,Turbet2018}, were stopped. \added{Note that the 1 and 10 bar pressures are estimated according to the planet's gravity. In our simulations, the integrated atmospheric mass column is scaled in $g_0/g$ with g the surface gravity of the planet and $g_0$ the surface gravity on Earth.} \\

\begin{table}[h!]
\renewcommand{\thetable}{\arabic{table}}
\caption{Parameters for  the TRAPPIST-1 planets in the HZ. $S_{\bigoplus}$, $M_{\bigoplus}$, $g_{\bigoplus}$ and $R_{\bigoplus}$ correspond to Earth's insolation, mass, gravity and radius, respectively. \added{The visibility corresponds to the maximum number of time the planets' transit will be observable during JWST 5 year nominal lifetime.}} \label{tab:TRAPPIST1}
\resizebox{0.75\textwidth}{!}{\begin{minipage}{\textwidth}
\begin{tabular}{c c c c}
\hline
\hline
Parameters & TRAPPIST-1e & TRAPPIST-1f & TRAPPIST-1g\\
\hline
Period (days) & 6.10 & 9.21 & 12.35 \\
Transit duration (s) & 3433  & 3756 & 4104 \\
$F_p$ ($S_{\bigoplus}$)& 0.662 & 0.382 & 0.258 \\
Mass ($M_{\bigoplus}$) & 0.772 & 0.934 & 1.148\\
Gravity ($g_{\bigoplus}$) & 0.930 & 0.853 & 0.871\\
Radius ($R_{\bigoplus}$) & 0.910 & 1.046 & 1.148\\
\added{Visibility (transits)} & 85 & 55 & 42\\
\hline
\hline
\end{tabular}
\end{minipage}}
\end{table}

\subsection{Simulation of the photochemistry}\label{sub:ATMOS}
Our 3-D model does not compute photochemistry prognostically.  Therefore, we use an off-line 1-D photochemistry code (Atmos) in order to compute the prevalence of minor gas species and organic hazes. To extend that set of gas species, a photochemical model has to be used in order to accurately simulate the formation and destruction of photochemical species and eventually, the formation of photochemical hazes. In this study, we used the Atmos 1-D model for our modern and Archean Earth-like simulations (\added{as neither sulfuric acid nor hydrocarbon hazes are expected to form in the CO$_2$-dominated atmospheres given our environmental assumptions}).\\  
 
Atmos is a 1-D radiative-convective climate model, coupled with a 1-D photochemistry model,  originally developed by James Kasting's group that has been used to determine the edges of the HZ, simulate an Archean Earth atmosphere, and study various exoplanets \citep{Arney2016,Arney2017,Lincowski2018,Meadows2018}. The 1-D representation of the atmosphere is plane-parallel at hydrostatic equilibrium.
 The vertical transport takes into account molecular and eddy diffusion. \added{Atmos includes molecules which have O, H, C, S, N and Cl atoms. For the modern Earth-like simulation, 309 reactions between 74 species have been considered while for the Archean Earth-like simulation, 459 reactions between 97 species have been used. Depending on the reaction, the JP-15 \citep{Burkholder2015} or the NIST 2005 (http://webbook.nist.gov/chemistry/fluid/) database was used.} The initial conditions such as the gas mixing ratios, out gassing fluxes and/or surface deposition velocities can be set at the top and bottom of the model. \added{Table \ref{tab:boundary} shows the boundary conditions used for the modern Earth-like and Archean Earth-like (from \cite{Charnay2013} case B) simulations. Boundary conditions from \cite{Lincowski2018} in Table 8 have been used for the modern Earth-like simulations, except for H$_2$O and clouds profiles, which have been provided from the LMD-G GCM outputs. For the Archean Earth-like simulation,  CH$_4$ mixing ratios have been fixed at the value used in the GCM. H$_2$, H$_2$S, SO$_2$ and CS$_2$ mixing ratio have also been fixed to help resolve convergence issues. Note that we have verified that fixing the mixing ratio of these species does not impact the transmission spectrum.  In both the modern Earth and Archean Earth -like simulations, the NO production from lightning \added{in the troposphere} is included at a rate of $1\times 10^9$ $molecules/cm^{2}/s$ (see NO flux in Table \ref{tab:boundary}). \deleted{For the modern Earth-like simulation only, sub-aerial emission of SO$_2$ at a rate of $9\times 10^9$ $molecules/cm^{2}/s$ up to 14~km has been included, along with a surface deposition velocity of $v=1$ $cm/s$. }}\\

\begin{table*}
\centering
\caption{Surface boundary conditions for the modern Earth-like and Archean Earth like (\cite{Charnay2013} case B) simulations. Table derived from \cite{Lincowski2018}. \added{VMR$_0$ corresponds to a fixed volume mixing ratio at the surface}, velocity depositions ($v$) are in units of $cm/s$ and fluxes ($F$) are in units of $molecules/cm^{2}/s$ \label{tab:boundary}}
\resizebox{0.8\textwidth}{!}{\begin{minipage}{\textwidth}
\centering
\begin{tabular}{c c c}
\hline
\hline
\added{Species} & \added{Modern Earth} & \added{Archean Earth (Charnay B)} \\
 & \added{boundary conditions} & \added{boundary conditions} \\
N$_2$ & VMR$_0$=0.78 & VMR$_0$=0.99\\
O$_2$  & VMR$_0$=0.21 & VMR$_0$=$4\times 10^{-8}$\\
CO$_2$  & VMR$_0$=$4\times 10^{-4}$ & VMR$_0$=$1\times 10^{-2}$\\
CH$_4$ & $F=6.8\times 10^{8}$ & VMR$_0$=$2\times 10^{-3}$\\
O & $v=1.0$ & $v=1.0$ \\
H & $v=1.0$ & $v=1.0$ \\
OH & $v=1.0$ & $v=1.0$ \\
HO$_2$ &  $v=1.0$ & $v=1.0$ \\
H$_2$O$_2$ &  $v=0.2$ & $v=1.0$ \\
H$_2$ & $v=2.4\times 10^{-4}$ & VMR$_0$=$5.3\times 10^{-7}$\\
CO & $v=0.03$, $F = 3.7 \times 10^{11}$& $v=1.2\times 10^{-4}$\\
HCO &$v=1.0$ & $v=1.0$ \\
H$_2$CO &$v=0.2$ & $v=0.2$\\
HNO & $v=1.0$ & $v=1.0$ \\
NO & $v=1.6\times 10^{-2}$, \replaced{$F = 1.0 \times 10^{9}$}{$F = 6.0 \times 10^{8}$} & $v=3.0\times 10^{-2}$,  $F = 1.0 \times 10^{9}$\\
NO$_2$ & $v=3.0\times 10^{-3}$ & $v=3.0\times 10^{-3}$\\
H$_2$S & $v=0.02$, $F = 2.0 \times 10^{8}$& VMR$_0$=$5.0\times 10^{-12}$\\
SO$_2$ & \deleted{$v=1.0$,} $F = 9.0 \times 10^{9}$   & VMR$_0$=$2.6\times 10^{-10}$ \\
H$_2$SO$_4$ & $v=1.0$, $F = 7.0 \times 10^{8}$ & $v=1.0$\\
HSO & $v=1.0$& $v=1.0$\\
OCS & $v=0.01$, $F = 1.5 \times 10^{7}$ & $v=0.01$\\
HNO$_3$ & $v=0.2$& $v=0.2$ \\
N$_2$O & \replaced{$F = 1.53 \times 10^{9}$}{v=0.0} & $v=0.0$\\
HO$_2$NO$_2$ &$v=0.2$ & $v=0.02$ \\
CS$_2$ &$F = 2.7\times 10^7$ & VMR$_0$=$3.0\times 10^{-11}$ \\
C$_2$H$_6$S & $F = 3.3\times 10^9$ & $F = 3.3\times 10^9$ \\
\hline
\end{tabular}
\end{minipage}}
\end{table*}

 The radiative transfer routine of Atmos uses the correlated-k absorption coefficients \citep{LacisOinas1991} derived from the HITRAN 2008 \citep{Rothman2009} and HITEMP 2010 \citep{Rothman2010} databases for pressures of $10^{-5}-10^2$ bar and for temperatures of 100-600~K. Photochemical haze (tholins) optical properties are derived from \cite{Khare1984} for the longwave and \cite{Gavilan2017} for the shortwave. The photochemically active wavelength range of the model recently has bee extended to now include the Lyman-$\alpha$ for a large set of species \citep{Lincowski2018}.\\
 
In this study, the photochemistry calculations are restricted to the terminator (longitude $\pm 90^\circ$) because it is the only region for which the atmosphere can be probed with transmission spectroscopy. Because photochemistry occurs on the substellar hemisphere, \added{ we assume here a solar zenith angle of $60^{\circ}$ and that the photochemical species are then transported toward the} terminator by dynamics \citep{Chen2018}. Note that the formation of hazes at lower zenith angles would not necessary increase the haze production rate because of the UV self-shielding by hazes \citep{Arney2016,Arney2017}. The transport of hazes from the day side to the terminator, requiring a full coupling between the GCM and photochemical model, is out of the scope of this paper but will be investigated in future studies.\\

To simulate the photochemical evolution around the terminator, we feed Atmos with the temperature/pressure profiles and mixing ratios from the LMD-G outputs for each latitude coordinate around the terminator. Because we use a $64\times 48$ longitude $\times$ latitude grid, Atmos is run 48 times around the terminator. To link the LMD-G GCM to the Atmos photochemical model, we have interpolated temperature and pressure profiles from the top of the GCM grid (going to $\sim 10^{-5}$~bar, i.e. about 65~km) to the top of the Atmos photochemical model grid (going up to $\sim0.05~Pa$, i.e about 100~km). \added{The temperature at the altitude of 100~km has been arbitrarily set to 150~K, similar to the thermosphere temperature on Earth. The temperature from the GCM lid to 100~km is then linearly decreasing to mimic the decreasing of temperature in Earth's mesosphere.} We made the following assumptions for these simulations:

\begin{itemize}
    \item Atmos is not coupled to LMD-G; we only feed mixing ratio and temperature / pressure profiles from LMD-G to Atmos and run the photochemical model.
   \deleted{ \item Initial conditions have been fixed by constant mixing ratios (from either modern or Archean Earth templates).}
    \item \added{No biomass fluxes are considered.}
    \item  Mixing ratios from LMD-G have been kept constant from the top of the GCM grid up to the top of the Atmos grid.
    \item The water profile modified by Atmos does not affect the clouds location and properties because the water photolysis appears in the upper atmosphere, beyond the upper limit of the GCM.
    \item Pressure and temperature profiles are extrapolated  from the top of the GCM grid up to the top of the Atmos grid. \added{Pressure decreases exponentially taking into account atmospheric scale height and the temperature decreases linearly down to 150~K at 100~km.}
    \item \added{The solar zenith angle (SZA) is fixed at 60$^{\circ}$, assuming that photochemical species at the terminator are produced and transported from the day side \citep{Chen2018}.}
\end{itemize}

Therefore, our methodology  is not a "coupling", and there is no feedback from the photochemical model to the GCM. A full coupling between LMD-G and Atmos will be investigated in future work.

When the photochemical model has converged, the new mixing ratios are computed for the following gases:  N$_2$, H$_2$O, CH$_4$, C$_2$H$_6$, CO$_2$, O$_2$, O$_3$, CO, H$_2$CO, HNO$_3$, NO$_2$, SO$_2$, N$_2$O and H$_2$, with some gases being more relevant either for the Modern Earth or the Archean Earth -like template. If aerosols (clouds and/or photochemical hazes) are formed, then atmospheric profiles of gas and aerosols are used to compute the transmittance  with PSG through the terminator for each of the TRAPPIST-1 planets in the HZ and each of the atmospheric configurations.

\subsection{Simulation of the transmission spectra}\label{sec:PSG}

We use the planetary Spectrum Generator (PSG, \cite{Villanueva2018}) to simulate JWST transmission spectra. PSG is an online radiative-transfer code for various objects of the solar system and beyond. PSG can compute planetary spectra (atmospheres and surfaces) for a wide range of wavelengths (UV/Vis/near-IR/IR/far-IR/THz/sub-mm/Radio) from any observatory, orbiter  or  lander and also includes a noise calculator.\\

\subsubsection{Aerosol optical properties}

\added{Four different kinds of aerosols are included in the simulated atmospheres: liquid and ice water, CO$_2$ ice, and fractal organic hazes. The optical properties of liquid and ice water as well as the fractal organic hazes are derived from HITRAN-RI 2016 from \cite{Massie2013}, while  CO$_2$ ice cloud optical properties are obtained from \cite{Hansen1991}.}\\

%%INCLUDE PARAGRAPH AND FIGURE ON OPTICAL PROPERTIES
%FIGURE 1
\begin{figure}[h!]
\centering
\resizebox{9cm}{!}{\includegraphics{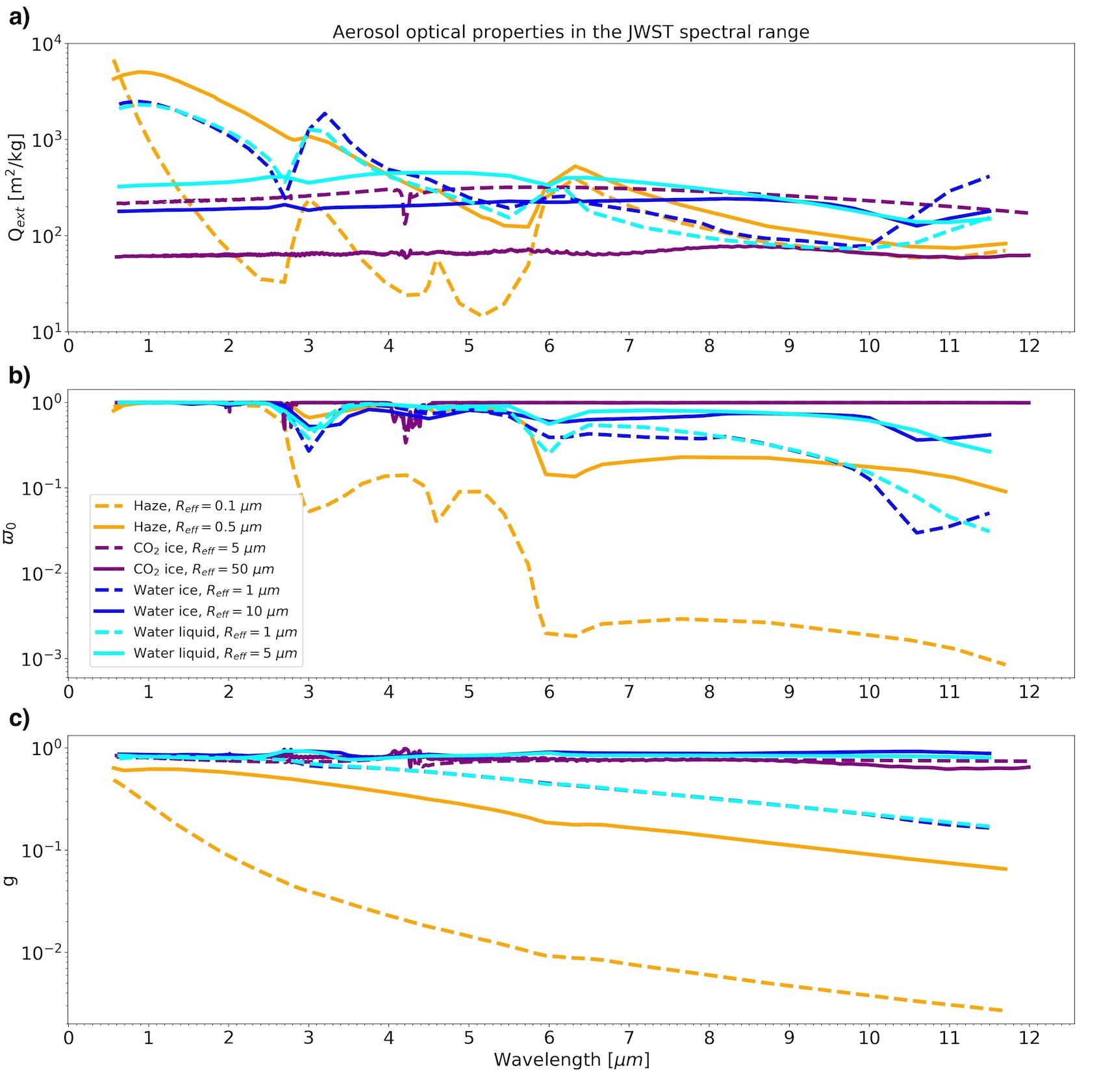}}
\caption{\replaced{From top to bottom:}{a)} extinction cross section (Q$_{ext}$), \added{b)} single scattering albedo ($\varpi_0$) and \added{c)} asymmetry parameter of the phase function (g) as a function of the wavelength across JWST NIRSpec Prism and MIRI LRS ranges.}
\label{fig:opticalprop}
\end{figure}

\added{Figure \ref{fig:opticalprop} shows the optical properties of the various aerosols formed by the LMD-G GCM (liquid water, ice water and CO$_2$ ice) and by the Atmos photochemical model (organic hazes). For each of them the dashed line represents the smallest effective radius (R$_{eff}$) and the solid line denotes the largest R$_{eff}$ formed in the atmosphere. \deleted{We can see that the smallest water droplets have a weak extinction cross section by comparison to the other aerosols, except for the hazes at R$_{eff}$ = 0.1 $\mu m$. } The single scattering albedo ($\varpi_0$) shows that the smallest hazes are strongly absorbing (small $\varpi_0$)  at most of the wavelengths while other aerosols produce a significant amount of scattering ($\varpi_0\ \sim $1) but progressively become more absorbent in the infrared. Similarly, the asymmetry parameter of the scattering phase function (g) shows that the scattering of the smallest hazes is almost isotropic (small g indicating a Rayleigh regime) while the other aerosols have g values close to 1 (forward scattering), progressively decreasing in the infrared for the smallest (1 $\mu m$) liquid and ice water particles. Therefore, this figure shows that the type of aerosol and their particle size have very different optical properties, which will have different impacts on the transmission of light through the planet's atmosphere.
}

\subsubsection{JWST instruments and noise}
\added{JWST has an aperture size of 6.5~m from tip to tip of its segmented mirror, which is equivalent to a 5.64~m diameter disk that we use in PSG.} We estimate that NIRSpec Prism is the most \added{suitable} instrument of JWST to characterize TRAPPIST-1 planets with transmission spectroscopy because it has a relatively wide wavelength coverage from 0.6 to 5.3 $\mu m$, for a resolving power (R) of 300 and TRAPPIST-1 should not reach the saturation of the detector. Note that partial saturation (in the SED peaks) strategy or alternate readout mode \citep{Batalha2018} can yield considerably better results in terms of fewer transits required to reach a desired S/N on gas detections  \citep{Batalha2018,Lustig_Yaeger2019}. 

In addition, some interesting gaseous features for the Earth-like atmospheres (Archean and modern) such as the  methane (CH$_4$) and ozone (O$_3$) absorption lines are also accessible  in the JWST MIRI \added{range. The mode supporting time series observation being the low}-resolution spectroscopy (LRS, R=100) with the wavelength range of $5.0-12.0\ \mu m$.  In this work, spectra are showed with R=300 across NIRSpec and MIRI ranges to \added{improve the visibility of the lines and} have continuous spectra from 0.6 to $20\ \mu m$ with a constant resolving power. \text{However, to calculate S/N in MIRI range we use R=100 or below.}\\

PSG includes a noise calculator to account for the following:  the noise introduced by the source itself (N$_{source}$), the background noise (N$_{back}$) following a Poisson distribution with fluctuations depending on $\sqrt{N}$ with  $N$  the mean number of photons received \citep{Zmuidzinas2003}, the noise of the detector (N$_{D}$) and the noise introduced by the telescope (N$_{optics}$). The total noise being then $N_{total}=\sqrt{N_{source}+N_{back}+N_D+N_{optics}}$.  

\section{Modern Earth-like atmospheres} \label{sec:Earth}
\subsection{Climate}
\begin{figure}[h!]
\centering
\resizebox{9cm}{!}{\includegraphics{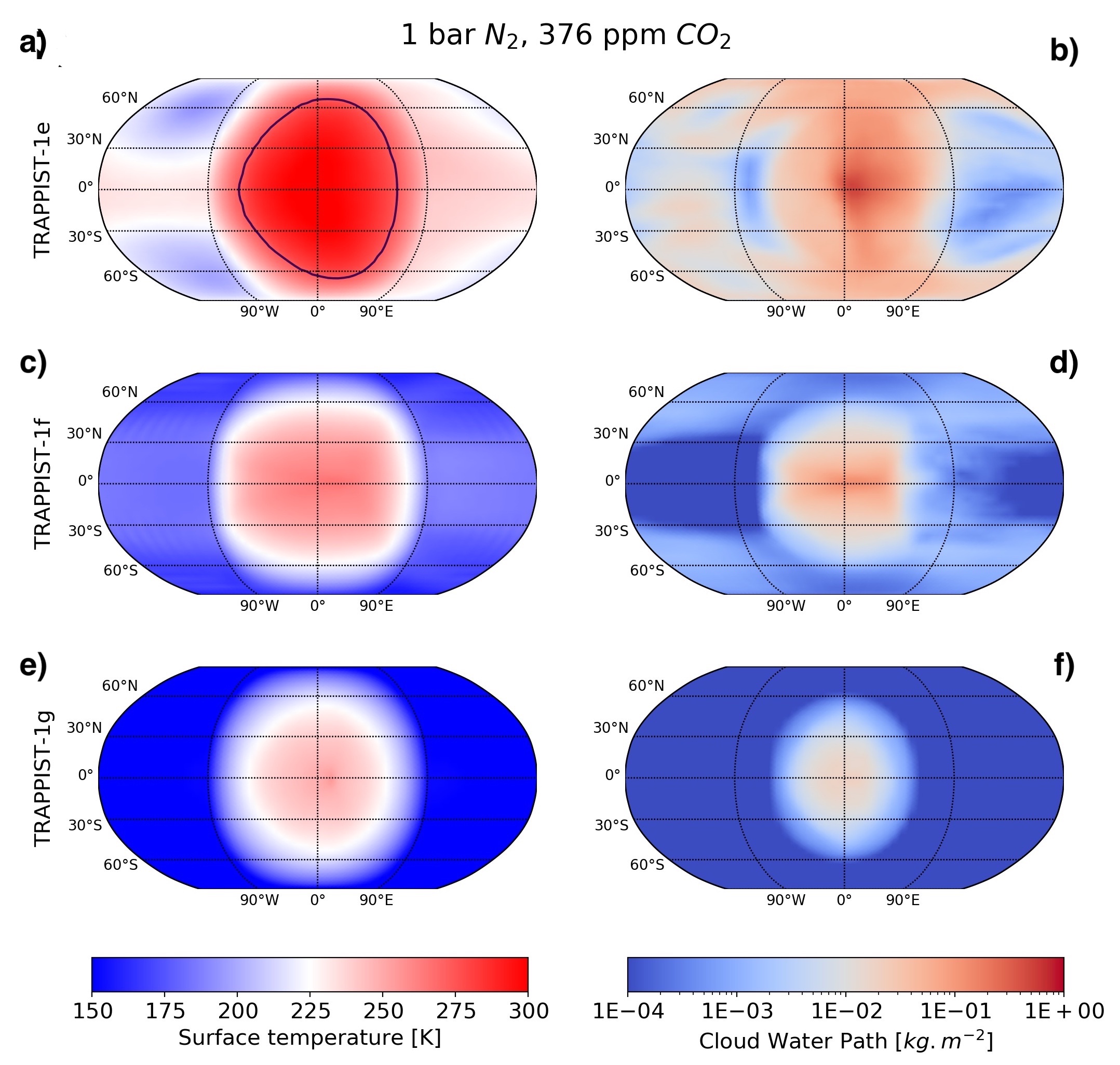}}
\caption{Surface temperature (left column) and cloud water path (right column) for TRAPPIST-1e (top row), 1f (middle row) and 1g (bottom row). The blue line shows the sea-ice boundary.  Note, only TRAPPIST-1e has any ice-fee ocean.}
\label{fig:modernClimates}
\end{figure}

\added{Because Earth is the only known inhabited planet and its habitability has been studied extensively,} when the question of a planet's habitability arises, \added{an atmosphere with boundary conditions based on the modern Earth as shown in Table \ref{tab:boundary} is always a key case to consider}. With LMD-G we have considered a 1 bar atmosphere composed of N$_2$ and 376~ppm of CO$_2$. While modern Earth consists of  $78\%$ of N$_2$ and $21\%$ of O$_2$ both gases have similar impacts on a planet's climate so we only take into account N$_2$ for the GCM simulations. O$_2$ and other minor gases will be considered in the photochemistry computation with Atmos. The surface temperature and water cloud path ($kg\cdot m^{-2}$) for TRAPPIST-1e, f and g are shown in Fig.~\ref{fig:modernClimates}. TRAPPIST-1e is the only planet to have an ice-free surface around the substellar point and a thick cloud deck going up to the terminator (especially at high latitudes pushed by Rossby waves from the substellar point). Thick clouds ($> 10^{-3}\ kg\cdot m^{-2})$ are present in the east terminator of TRAPPIST-1f but not in TRAPPIST-1g. Table~\ref{tab:climateModern} lists the mean, maximum and minimum surface temperature as well as the integrated column of condensed species for TRAPPIST-1e, f and g. We can see that both the surface temperatures and amount of condensed species are much lower for the modern Earth-like atmosphere than for the CO$_2$ dominated atmosphere at 1 bar surface pressure (see Table~\ref{tab:climateCO2}). The mean surface temperature of TRAPPIST-1e (244~K) is in very good agreement with 3-D climate simulations with CAM4 GCM \cite{Wolf2017} (241~K) and LMD-G GCM \cite{Turbet2018} (248~K; 4~K difference arises due to the use of updated planetary parameters) but far from the 1-D simulation of \cite{Lincowski2018} (279~K and 282~K for the clear and cloudy simulations, respectively).
%FIGURE 2

\begin{table}
\centering
\caption{Surface temperatures (TS) and integrated column of condensed species for the modern Earth-like atmosphere. Values with an asterisk are averaged around the terminator only.} \label{tab:climateModern}
\resizebox{0.75\textwidth}{!}{\begin{minipage}{\textwidth}
\begin{tabular}{c| c  | c | c }
\hline
\hline
\added{Parameters} & \multicolumn{3}{c}{Planets}\\
  & \added{TRAPPIST-1e} & \added{TRAPPIST-1f} & \added{TRAPPIST-1g}\\
TS mean [$K$] & 244 &  197 & 168  \\
TS min [$K$] & 194 & 157 &   126   \\
TS max [$K$]  &  304 & 266  &  256  \\
H$_2$O liq* [$10^{-3}kg\cdot m^{-2}$] & 1.3 & 0.0 & 0.0   \\
H$_2$O ice* [$10^{-3}kg\cdot m^{-2}$] & 13.6 & $6.0\times 10^{-1}$ & $1.3\times 10^{-2}$  \\
CO$_2$ ice* [$10^{-3}kg\cdot m^{-2}$] & 0.0 & 0.0  &  0.0 \\
\hline
\hline
\end{tabular}
\end{minipage}}
\end{table}

%FIGURE 3
\begin{figure}[h!]
\centering
\resizebox{9cm}{!}{\includegraphics{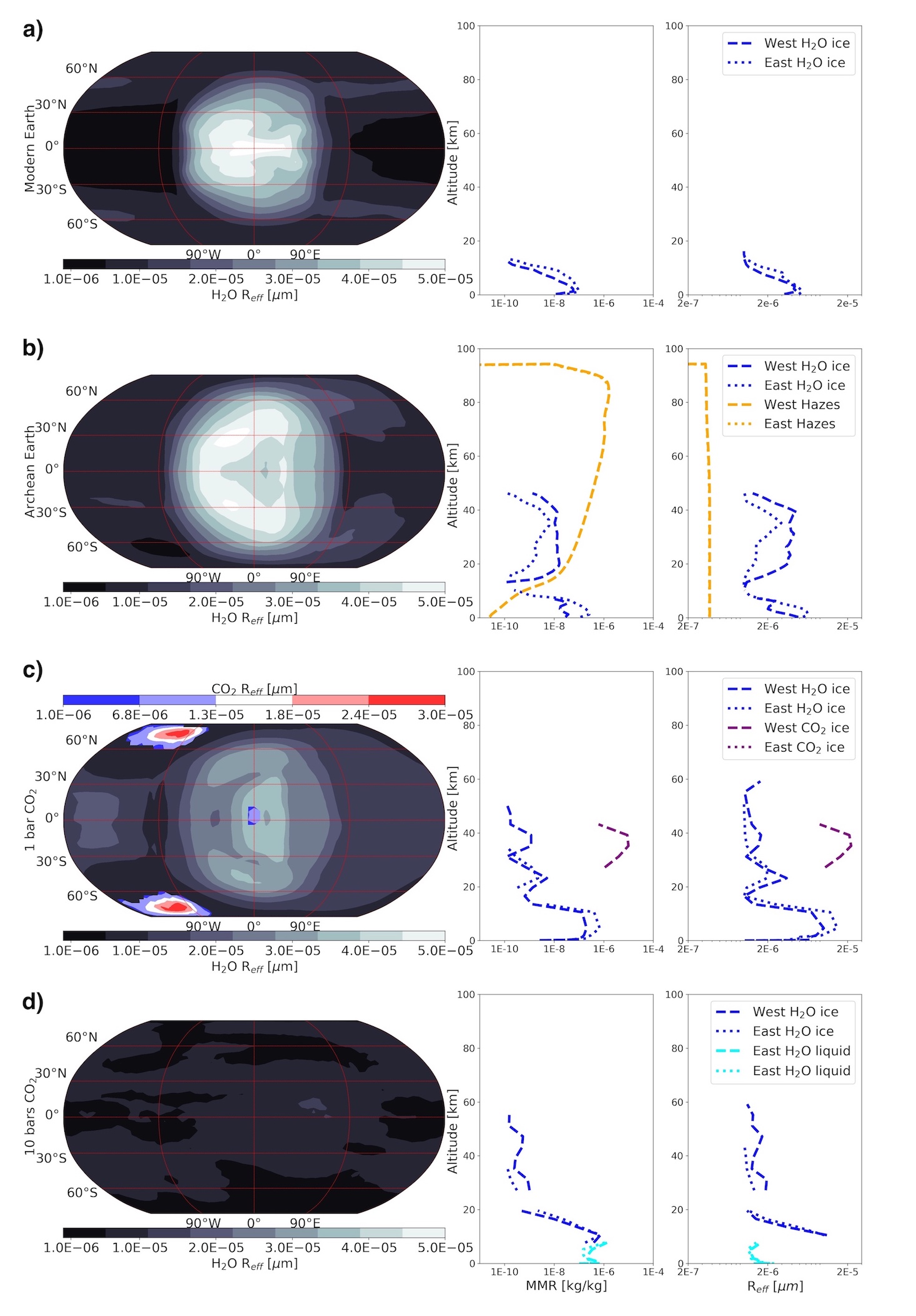}}
\caption{Left panel: vertically averaged cloud effective radius (R$_{eff}$), right panel: vertical distribution of the mass mixing ratio (MMR) and R$_{eff}$ at the East and West terminator for various aerosols \added{in the TRAPPIST-1f atmosphere}. Note that hazes are only generated with the Atmos photochemical model. When the MMR is very low, typically below $10^{-10}$ kg/kg, the R$_{eff}$ is tiny and its dimension and optical properties are poorly constrained. LMD-G set up therefore the R$_{eff} =10^{-6}\ \mu m$ threshold as the minimum value to be output. Therefore,  MMR value below $10^{-10}$  kg/kg and R$_{eff}\ \leq 10^{-6}\ \mu m$ have been excluded from the averaging.}
\label{fig:Reff}
\end{figure}

\added{Fig. \ref{fig:Reff} shows the vertically-averaged horizontal distribution of the cloud effective radius R$_{eff}$ and their vertical distribution at the terminator along with the mass mixing ratio (MMR) for TRAPPIST-1f in the four atmospheric scenarios considered in this study: modern Earth, Archean Earth, 1 and 10~bars CO$_2$-dominated atmospheres. The largest R$_{eff}$ are associated with the largest MMR due to the coalescence processes.  The atmosphere with boundary conditions based on the modern Earth (see Table \ref{tab:boundary})  is the coldest of the four scenarios and we can see that clouds are confined closer to the surface than in the  other three scenarios and will therefore have a smaller impact on the transmission spectra.  Also, CO$_2$ ice and hazes form at higher altitudes than H$_2$O clouds and with larger MMR. Therefore, we  expect a stronger impact from them on the spectra than H$_2$O clouds. As in Fig. \ref{fig:modernClimates}, we can see an asymmetry between the East and West side of the substellar point where most of the water clouds are advected \deleted{West side of the substellar point,} toward the East terminator because of Kelvin waves introduced by the Coriolis force. Therefore, there are more clouds (larger MMR) and larger R$_{eff}$ at the East terminator than the West terminator. Such a difference can be potentially detected with time resolved observations of the ingress and egress \citep{Line2016} but this is out of the scope of this study. Discussions about the three other scenarios in Fig \ref{fig:Reff} will be presented in the corresponding sections.}\\

\begin{figure}[h!]
\centering
\resizebox{9cm}{!}{\includegraphics{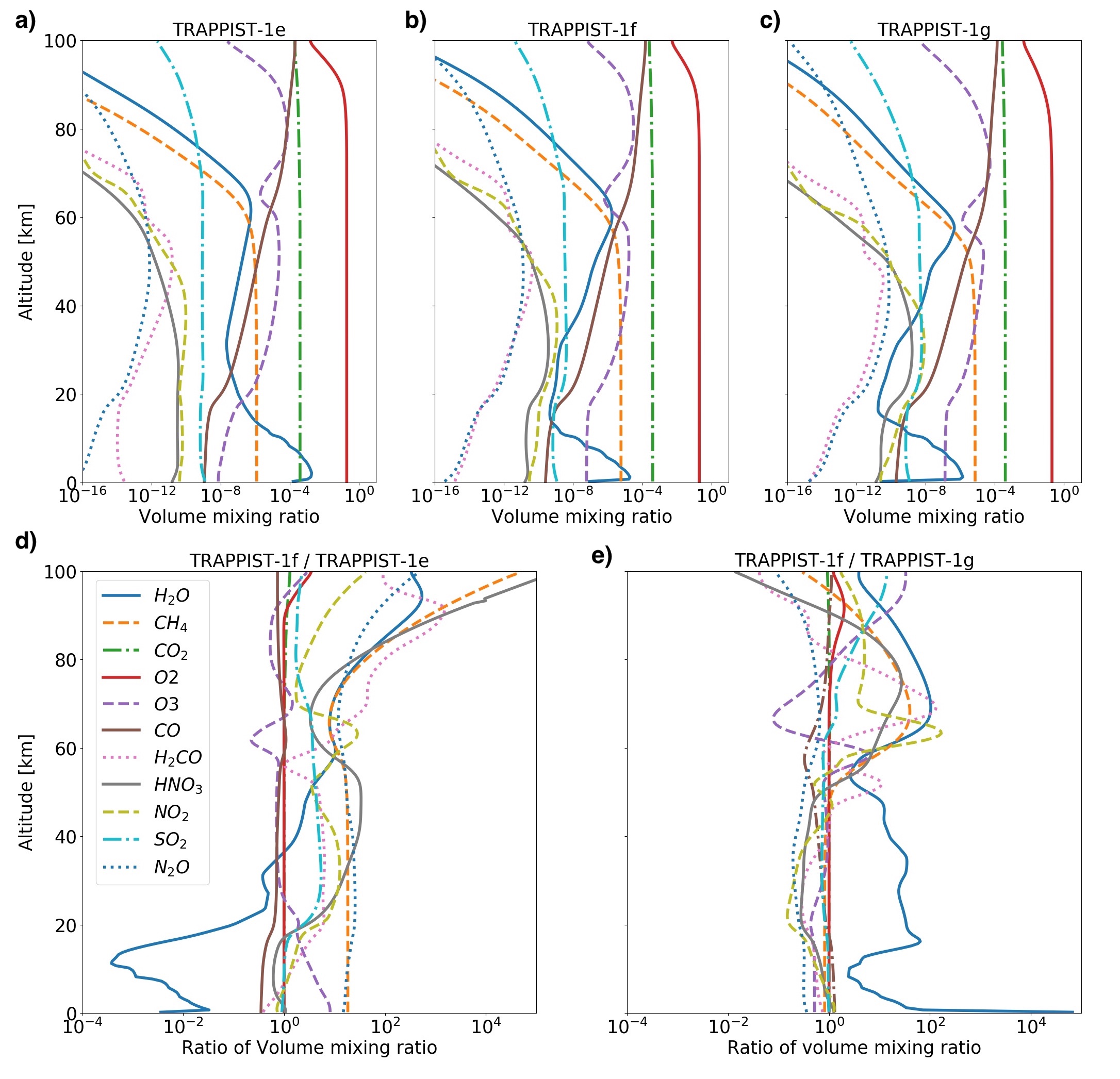}}
\caption{Gas mixing ratio profiles \added{for atmospheres with boundary conditions based on the modern Earth (see Table \ref{tab:boundary})} for \added{a)} TRAPPIST-1e,  \added{b)} 1f and  \added{c)} 1g, as well as profile ratios between  \added{d)} TRAPPIST-1f and 1e and  between \added{e)} TRAPPIST-1f and 1g .}
\label{fig:modernProfiles}
\end{figure}

To simulate more gases in the planet's atmosphere than the ones used in the GCM climate module (N$_2$, CO$_2$, CH$_4$ and H$_2$O), we run the Atmos photochemical model at the terminator (ignoring feedback on the climate) from atmospheric profiles (temperature, pressure and mixing ratios) computed by LMD-G. Figure \ref{fig:modernProfiles} shows a selection of averaged atmospheric profiles around the terminator for TRAPPIST-1e, 1f and 1g obtained with Atmos \added{and the atmospheric profile ratios between TRAPPIST-1f and 1e and between TRAPPIST-1f and 1g. There is a significant amount of H$_2$O in the lower atmosphere of TRAPPIST-1e compared to TRAPPIST-1f, because TRAPPIST-1e is warmer, and hence  more water is brought into the atmosphere from convection and evaporation on the day side which is then transported to the terminator. However, this trend is reversed  in the upper atmosphere (above $\sim40$~km) where H$_2$O in TRAPPIST-1e atmosphere, closer to the star, is strongly photodissociated. We can also see that CH$_4$ is about 70 times more abundant in TRAPPIST-1f below 60~km where the reaction CH$_4$ + OH $\rightarrow$ CH$_3$ + H$_2$O is two order of magnitudes faster in TRAPPIST-1e (the excess of OH being produced by the larger amount of H$_2$O from the reaction H$_2$O + O$_1$D  $\rightarrow$   OH + OH);  it then increases by a few orders of magnitude near TOA. \replaced{TRAPPIST-1f/TRAPPIST-1e ratios of O$_3$ and NO$_2$ show an anti-correlation. M-dwarfs generally produce less radiation than the Sun in the 200-350~nm range where O$_3$ is photolysed; therefore O$_3$ abundances are significantly higher than for Earth around Sun. For TRAPPIST-1e, more free oxygen is available from water photolysis to react with NO$_2$ than in TRAPPIST-1f; therefore they are not available to form O$_3$. This pathway is explained in \cite{Harman2018} Table 1, cycle 4. Note that the volume mixing ratio (VMR) of NO$_2$ is so small that its spectral signature will not be observable in the spectra.}{Ozone responses are complex depending on UV availability and trace species like HOx and NOx \citep{Grenfell2014, Harman2018}}. Concerning the N$_2$O, its reactivity does not seem significant in the lower atmosphere with no change in VMR. Above $\sim60$~km, the photodissociation of N$_2$O by UVB \citep{Segura2003,Grenfell2014} becomes important, reducing its relative proportion in TRAPPIST-1e.  Finally, O$_2$ does not appear to be too different between TRAPPIST-1 1e and 1f, except in the upper atmosphere ($\geq90$~km) where O$_2$ is more photolyzed for TRAPPIST-1e, closer to the star. Between TRAPPIST-1f and 1g, 1f receives the more flux from TRAPPIST-1 and has the wetter atmosphere leading to less CH$_4$, O$_3$  and N$_2$O overall but more NO$_2$. But, near TOA, both O$_2$ and O$_3$ build up easier for TRAPPIST-1f because more H$_2$O is photodissociated leading to more free oxygen.} The atmospheric profiles presented here are then used, along with cloud profiles from LMD-G,  to simulate the transmission spectra.  
 
 %FIGURE 4

\subsection{JWST simulated spectra: Impact of H$_2$O clouds}
\begin{figure}[h!]
\centering
\resizebox{9cm}{!}{\includegraphics{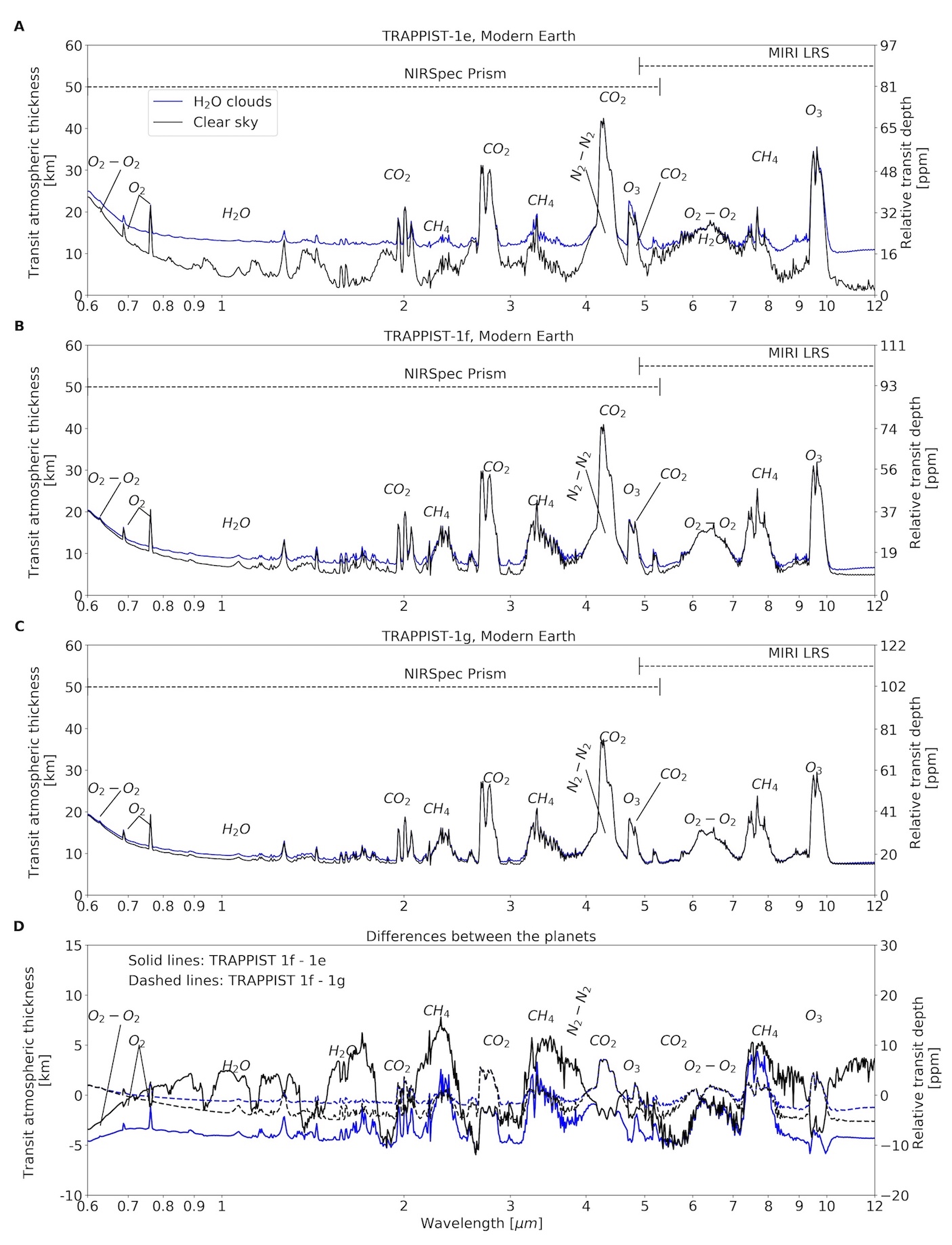}}
\caption{Simulated transmission spectra by JWST NIRSpec Prism and MIRI with R=300 for aquaplanets A) TRAPPIST-1e, B) 1f  and C) 1g \added{with boundary conditions based on the modern Earth as shown in Table 2). Panel D) shows  differences between planetary spectra.}}
\label{fig:modernSpectra}
\end{figure}

The NIRSpec Prism and MIRI transmission spectra at R=300 for TRAPPIST-1e, f and g with the with boundary conditions based on the modern Earth  are presented in Fig.~\ref{fig:modernSpectra}. The relative transit depth, the signal-to-noise ratio  for 1 transit (S/N-1) and the number of transits needed to achieve $3\ \sigma$ and a $5\ \sigma$ detections are summarized in Table~\ref{tab:transitmodern} for selected absorption lines. A resolving power of R=30  has been adopted to optimize the S/N \citep{Morley2017}  while determining the number of transits.

The mathematical expression between the relative transit depth and the transit atmospheric thickness is \citep{Winn2010}:

\begin{eqnarray}
\label{eq:transit}
\begin{split}
\delta\Delta=(\delta R / R_s)^2 + (2 \times R_p \times \delta R)/R_s^2\\
\sim (2 \times R_p \times \delta R)/R_s^2
\end{split}
\end{eqnarray}

with $\Delta$ and $\delta\Delta$ the transit depth and relative transit depth, respectively, in part-per-million, $R_p$ the planet's radius, $\delta R$ the transit atmospheric thickness and $R_s$ the radius of the star in kilometer unit. Note that $(R_p/R_s)^2$ represents the transit depth for an air-less planet. We can see that $\delta\Delta$ is dependent on the planet's radius. The planetary radius is increasing from TRAPPIST-1e to -1f and -1g (see Table \ref{tab:TRAPPIST1}), \added{leading to the largest} transit depth for TRAPPIST-1g. Note that $\delta\Delta$ and  $\delta R$ decrease with decreasing temperature or increasing gravity. Yet, TRAPPIST-1g is the coldest planet but also one with the lowest gravity. \added{Furthermore, the atmospheric refraction increases with the planet's distance to the host star leading to an increase in the altitude of the continuum and therefore a reduction of the relative transit depth of the lines. Finally, Table \ref{tab:TRAPPIST1} also reports the number of times the planets will transit in front of TRAPPIST-1 when being in the visibility zone of JWST during its nominal lifetime of 5 years. This number is reduced as the orbital period increases (85, 55 and 42 transits for TRAPPIST-1e, 1f and 1g, respectively).} \\

%FIGURE 5

\begin{table*}
\centering
\caption{Relative transit depth (ppm), signal-to-noise ratio  for 1 transit (S/N-1) and number of transits to achieve  $5\ \sigma$ and $3\ \sigma$ detection for various spectral lines of the modern Earth-like atmosphere. Numbers in parentheses are for clear sky only while numbers without parentheses are the real values accounting for the impact of clouds. The hyphen represents the cases for which more than 100 integrated transits are needed \added{and the * mark denotes the values above the maximum number of transits per planet during JWST nominal lifetime mentioned in Table \ref{tab:TRAPPIST1}.}} \label{tab:transitmodern}
\resizebox{0.8\textwidth}{!}{\begin{minipage}{\textwidth}
\centering
\begin{tabular}{c c c  c }
\hline
\hline
 Planets & \added{TRAPPIST-1e} & \added{TRAPPIST-1f} & \added{TRAPPIST-1g}\\
\hline
Instrument & \multicolumn{3}{c}{\added{NIRSpec Prism (R=30)}} \\
\hline
Feature & \multicolumn{3}{c}{O$_2$ $0.8\ \mu m$}\\
Depth [ppm] & 5(10) & 9(10) & 10(11)  \\
S/N-1  &  0.0(0.1) & 0.1(0.1) & 0.1(0.1)  \\
N transits  ($5\sigma$) & -(-) & -(-) & -(-)    \\
N transits  ($3\sigma$) & -(-) & -(-) & -(-)    \\
\hline
Feature & \multicolumn{3}{c}{H$_2$O $1.4 \ \mu m$} \\
Depth [ppm] & 2(10) & 2(3) & 3(3)  \\
S/N-1  &  0.0(0.2) & 0.1(0.1) & 0.1(0.1)  \\
N transits  ($5\sigma$) & -(-) & -(-) & -(-)    \\
N transits  ($3\sigma$) & -(-) & -(-) & -(-)    \\
\hline
Feature & \multicolumn{3}{c}{CH$_4$ $3.3\ \mu m$}  \\
Depth [ppm] & 12(20) & 24(27) & 25(26)    \\
S/N-1  &  0.2(0.5) & 0.5(0.6) & 0.5(0.7) \\
N transits  ($5\sigma$) & -(-) & -(76*) & 83*(74*)  \\
N transits  ($3\sigma$) & -(55) & 36(27) & 30(27)  \\
\hline
Feature  & \multicolumn{3}{c}{CO$_2$ $4.3\ \mu m$} \\
Depth [ppm] & 47(61) & 60(63) & 58(60)   \\
S/N-1  & 0.8(1.1)  & 1.1(1.2) & 1.2(1.2)  \\
N transits  ($5\sigma$) & 35(21) & 20(18) & 19(18) \\
N transits  ($3\sigma$) & 13(8) & 7(6) & 7(6) \\
%\hline
%Feature  & \multicolumn{3}{c}{$N_2-N_2$ $4.3\ \mu m$} \\
%Depth [ppm] & 11(26) & 20(29) & 29(31)   \\
%S/N-1  & 0.17(0.41)  & 0.33(0.48) & 0.51(0.54)  \\
%N transits  ($5\sigma$) & 54(35) & 39(32) & 39(31) \\

\hline
Instrument &\multicolumn{3}{c}{\added{MIRI MRS (R=30)}} \\
\hline
Feature & \multicolumn{3}{c}{$O_2-O_2$ $6.5\ \mu m$} \\
Depth [ppm] & 14(27) & 22(26) & 23(25)    \\
S/N-1  &  0.1(0.2) & 0.2(0.2) & 0.2(0.2)   \\
N transits  ($5\sigma$) & -(-) & -(-) & -(-)   \\
N transits  ($3\sigma$) & -(-) & -(-) & -(-)   \\
\hline
Feature & \multicolumn{3}{c}{CH$_4$ $7.7\ \mu m$}  \\
Depth [ppm] & 13(24) & 29(33) & 31(32)    \\
S/N-1  &  0.1(0.2) & 0.2(0.2) & 0.2(0.2)   \\
N transits  ($5\sigma$) & -(-) & -(-) & -(-)   \\
N transits  ($3\sigma$) & -(-) & -(-) & -(-)   \\
\hline
Feature & \multicolumn{3}{c}{O$_3$ $9.6\ \mu m$} \\
Depth [ppm] & 36(48) & 43(47) & 44(46)    \\
S/N-1  &  0.1(0.2) & 0.2(0.3) & 0.2(0.4)   \\
N transits  ($5\sigma$) & -(-) & -(-) & -(-)   \\
N transits  ($3\sigma$) & -(-) & -(-) & -(-)   \\

\hline
\hline
\end{tabular}
\end{minipage}}
\end{table*}

We can see \added{in Fig. \ref{fig:modernSpectra} for TRAPPIST-1e} that H$_2$O clouds raise the continuum level up to a few kilometers above the surface, flattening the H$_2$O lines and reducing the relative transit depth (or atmospheric thickness) of other species. \added{TRAPPIST-1f and 1g are much less affected by clouds because the weakest convection farther away from the star mutes the cloud formation.}
We have determined that the H$_2$O line at $1.14\ \mu m$ is the strongest H$_2$O line not being blended by CO$_2$ for such an atmosphere. Indeed, even the well-known $2.7\ \mu m$ H$_2$O line is completely dominated by CO$_2$ in this same spectral region, because H$_2$O is confined to the lower atmosphere where the opacity to the infrared radiation is high and where clouds are located. However, the relative transit depth of that H$_2$O line, or any other, is so low (only a few ppm) that it is very challenging to detect.\\

\added{In Fig. \ref{fig:modernSpectra} bottom subplot, we can see that the differences between the planets is the largest between TRAPPIST-1e and 1f (solid lines), transiting from a wet and cloudy atmosphere to a drier and mostly cloud-free atmosphere. The largest difference between the clear sky spectra concerns CH$_4$, which is strongly muted in TRAPPIST-1e where it is destroyed by OH produced by the large amount of H$_2$O. TRAPPIST-1f and 1g are similar, both in a snowball state (see Fig. \ref{fig:modernClimates}), and the difference between their transmission spectra (dash lines) is small.}

 \added{In the NIRSpec prism range, only CO$_2$ at $4.3\ \mu m$ is detectable at $5\ \sigma$ during JWST's nominal lifetime (TRAPPIST-1e, 1f and 1g transiting 85, 55 and 42 times, respectively). Note that 35 and 21 transits are required to detect CO$_2$ at $5\ \sigma$ for the cloudy and clear sky TRAPPIST-1e, respectively. These results are in relatively good agreement with \cite{Lustig_Yaeger2019} transit values obtained with NIRSpec Prism.  CH$_4$ at $3.3\ \mu m$ could be detectable for TRAPPIST-1f and 1g at $3\ \sigma$ but the transit depth of about 25~ppm could be below the noise floor (see section \ref{sec:discussion}). In the MIRI range, while some features like O$_3$ at $9.6\ \mu m$ offer transit depths of the order of 40~ppm, the larger noise does not allow any detection at 3 or $5\ \sigma$ in less than 100 transits.}\\

\section{Archean Earth-like atmospheres} \label{sec:Archean}
\subsection{Climate}
The climate of the Archean era (3.8-2.5 Ga) is still being debated. In this study we chose to use the three Archean Earth atmospheric compositions by \cite{Charnay2013} that were previously simulated with LMD-G. Those configurations for a 1 bar surface pressure are dominated by N$_2$ with the following amount of GHG:
\begin{itemize}
    \item Charnay case A (900~ppm of CO$_2$, 900~ppm of CH$_4$)
    \item Charnay case B (10,000~ppm of CO$_2$, 2,000~ppm of CH$_4$)
    \item Charnay case C (100,000~ppm of CO$_2$, 2,000~ppm of CH$_4$)
\end{itemize}

\begin{figure*}[h!]
\centering
\resizebox{12cm}{!}{\includegraphics{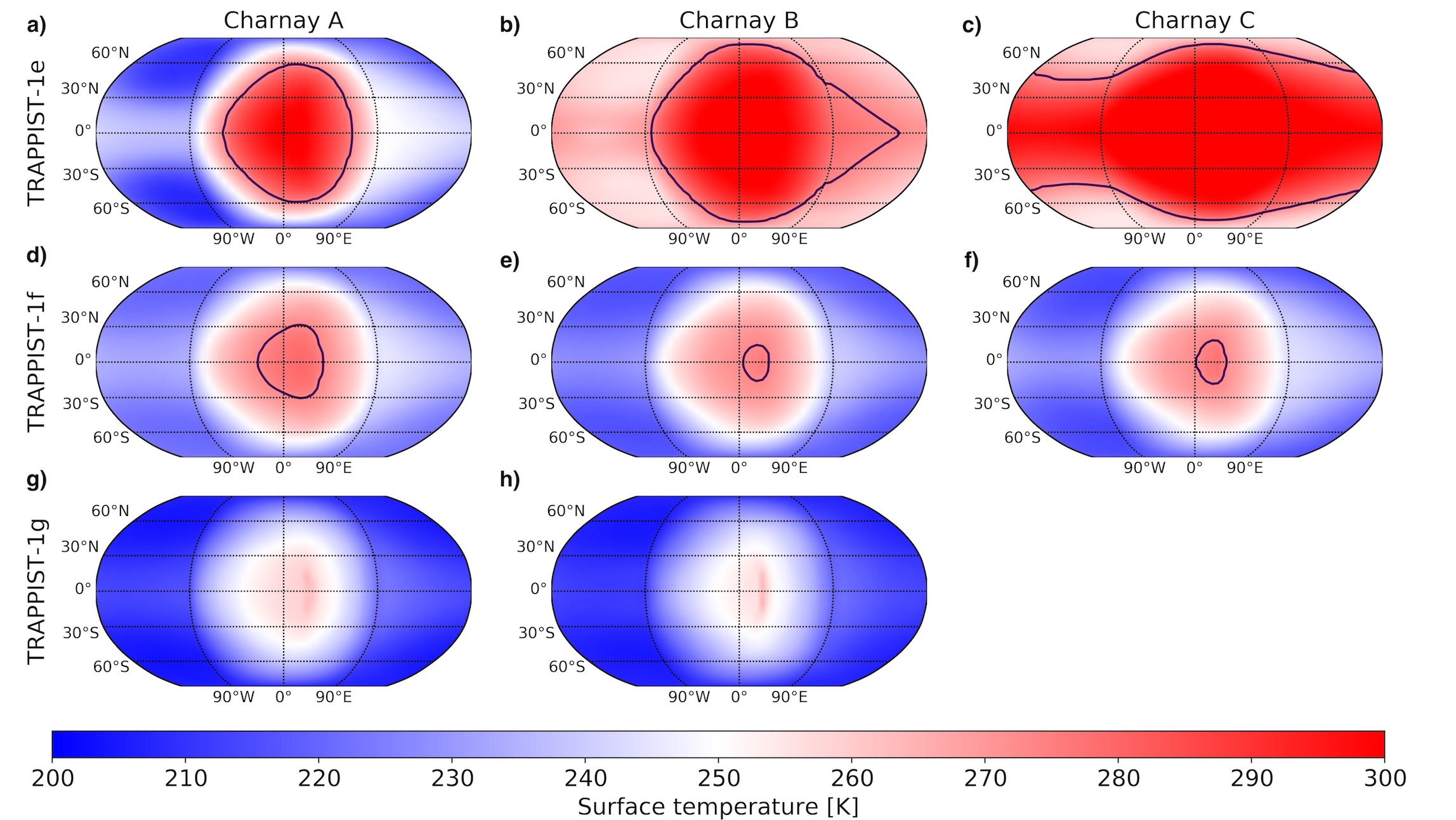}}
\caption{Surface temperature map in Kelvin (K) for aquaplanets TRAPPIST-1e (top row), TRAPPIST-1f (middle row) and TRAPPIST-1g (bottom row) for \cite{Charnay2013} case A (left), case B(middle) and case C (right) Archean Earth atmospheres. TRAPPIST-1g case C is missing because CO$_2$ in the atmosphere has condensed to the night side leading to the crash of the simulation. The blue line shows the sea-ice boundary.}
\label{fig:ArcheanSurfTemp}
\end{figure*}

\begin{figure*}[h!]
\centering
\resizebox{12cm}{!}{\includegraphics{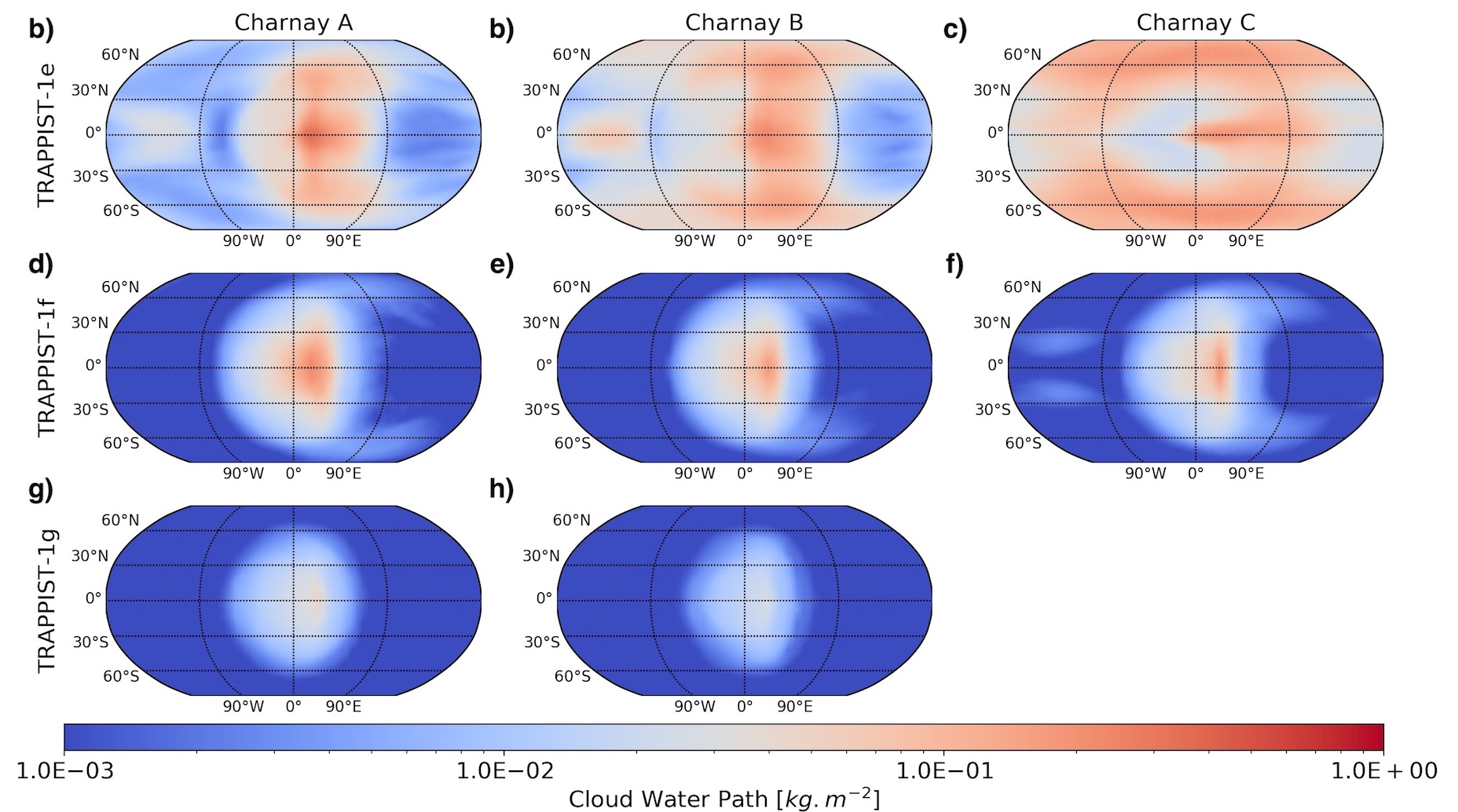}}
\caption{Integrated cloud water column in $kg.m^{-2}$ for aquaplanets TRAPPIST-1e (top row), 1f (middle row) and 1g (bottom row) for \cite{Charnay2013} case A (left), case B (middle) and case C (right) Archean Earth atmospheres. TRAPPIST-1g case C is missing because CO$_2$ in the atmosphere has condensed to the night side leading to the crash of the simulation.}
\label{fig:ArcheanH2O}
\end{figure*}

The surface temperatures and the water cloud columns are displayed in Fig. \ref{fig:ArcheanSurfTemp} and Fig. \ref{fig:ArcheanH2O}, respectively. Mean, \added{minimum and maximum} values  are reported in Table \ref{tab:climateArchean}.\\ Again, we were not able to find a stable climate state for TRAPPIST-1g with a Charnay case C atmosphere, because CO$_2$ condenses on the night side. \\
In Fig. \ref{fig:ArcheanSurfTemp} we can see that the surface temperature is increasing from Charnay case A to Charnay case C for TRAPPIST-1e while for TRAPPIST-1f and -1g, the surface temperature is maximum for Charnay case A, followed by case C and finally case B. On the one hand, for TRAPPIST-1e, the atmosphere is warm and moist and the water feedback has a large effect, as well as the change of albedo due to clouds and the ratio of water/ice surfaces; on the other hand for TRAPPIST-1f and -1g, their dryer atmosphere leads to a weak water feedback and the increase of CH$_4$ from Charnay Case A to Charnay case B promotes an anti-greenhouse effect, more powerful than the increase of CO$_2$. As a result, Charnay case B is the coolest. In  Fig. \ref{fig:ArcheanH2O} the relative amount of condensed water between the cases follows the surface temperature, with the largest cloud coverage for TRAPPIST-1e being Charnay case C, while for TRAPPIST-1f and -1g it is Charnay case A.\\

%FIGURE 6

%FIGURE 7

\begin{table}
\centering
\caption{Surface temperatures (TS) and integrated column of condensed species for the Archean Earth-like atmosphere. Values with an asterisk are averaged around the terminator only.} \label{tab:climateArchean}
\resizebox{0.75\textwidth}{!}{\begin{minipage}{\textwidth}
\begin{tabular}{c| c c c  | c c c | c c }
\hline
\hline
\added{Parameters} & \multicolumn{8}{c}{Planets}\\
  & \multicolumn{3}{c}{\added{TRAPPIST-1e}} & \multicolumn{3}{c}{\added{TRAPPIST-1f}} & \multicolumn{2}{c}{\added{TRAPPIST-1g}}\\
\added{Case} & A & B & C & A & B & C & A & B\\
TS mean [$K$] & 243 & 273& 286 & 238 & 234& 235 & 204& 221\\
TS min [$K$] &207 & 254 & 254 & 219 & 215 & 214 & 203 & 205 \\
TS max [$K$]  & 305 & 310 & 324 & 279 & 277 & 278 & 204 &  264 \\
H$_2$O liq* [$10^{-3}kg\cdot m^{-2}$] & 3.7 & 12.0 & 74.5 & 0.0 & 0.0 & 0.0 & 0.0 &  0.0 \\
H$_2$O ice* [$10^{-3}kg\cdot m^{-2}$] &12.6 & 24.7  & 26.1 & 1.1& 0.81& 0.93 & 0.07 &  0.05 \\
\hline
\hline
\end{tabular}
\end{minipage}}
\end{table}

\begin{figure}[h!]
\centering
\resizebox{8cm}{!}{\includegraphics{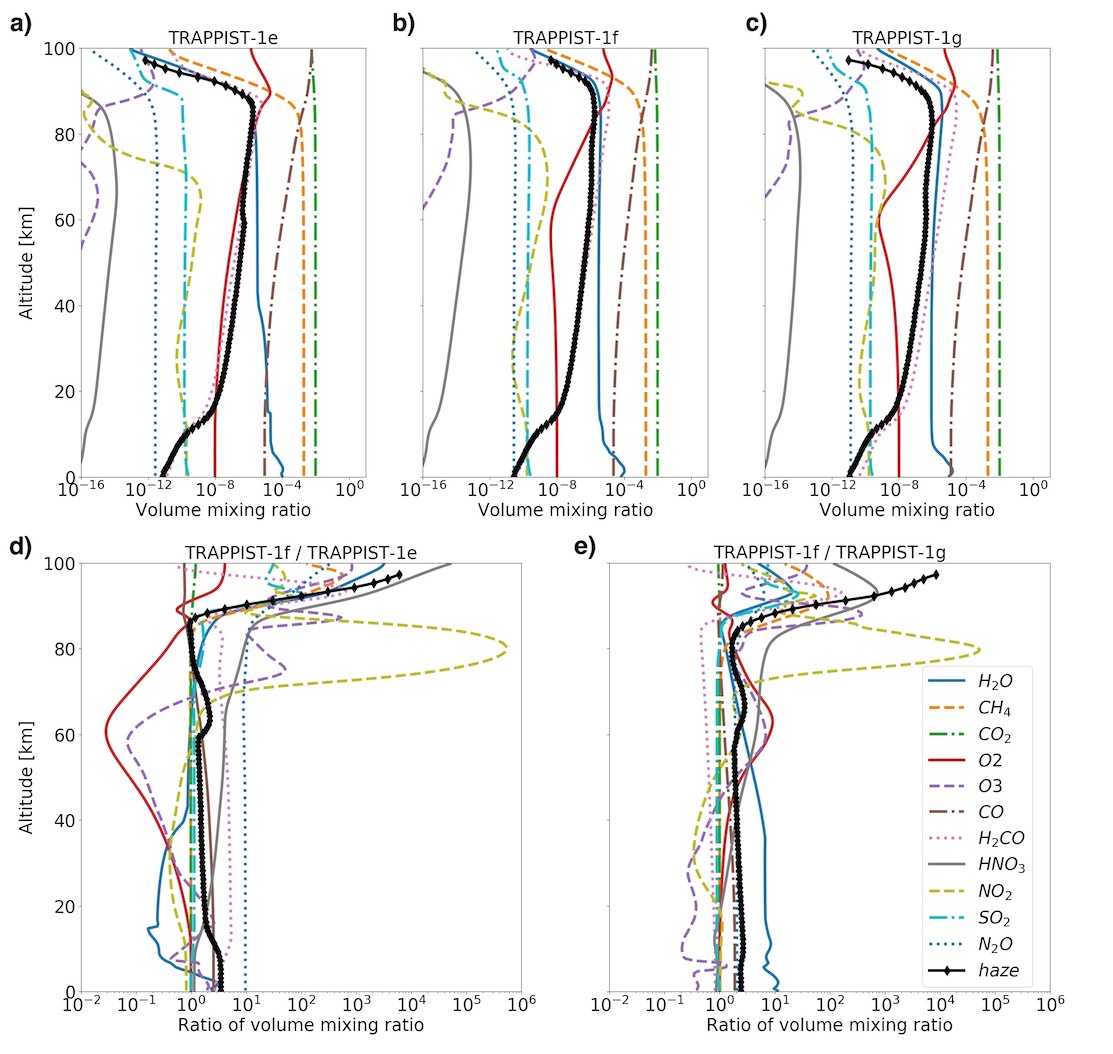}}
\caption{Gas mixing ratio profiles for an Archean Earth-like composition of \cite{Charnay2013} case B for \added{a)} TRAPPIST-1e,  \added{b)} 1f and  \added{c)} 1g, as well as profile ratios between  \added{d)} TRAPPIST-1f and 1e and  between \added{e)} TRAPPIST-1f and 1g.}
\label{fig:ArcheanProfiles}
\end{figure}

For planets for which the ratio of methane over carbon dioxide  (CH$_4$/CO$_2$) in the atmosphere exceeds about 0.1, haze formation can occur \citep{Arney2016}. Such hydrocarbon haze is generated by methane photolysis from $Ly\ \alpha$. Only Charnay Case A and B have the required CH$_4$/CO$_2$ to produce photochemical hazes. For this study, we have performed the photochemistry and transmission spectra simulations only for Charnay case B. This case offers larger concentrations of CO$_2$ and CH$_4$ than Charnay case A and can produce photochemical hazes contrary to Charnay case C. The Charnay case B Archean Earth-like atmospheric profiles is shown in Fig. \ref{fig:ArcheanProfiles}.  \added{In the Archean Earth-like GCM simulations, since fixed mixing ratios of CO$_2$, CH$_4$ and N$_2$ are used, they are also  constants for the photochemistry simulations with Atmos, along with other gases such as O$_2$, H$_2$, H$_2$S, SO$_2$ and C$_2$H$_6$S. These fixed boundary conditions lead to no major differences in the lower atmosphere between TRAPPIST-1e, 1f and 1g profiles.
However, TRAPPIST-1e receives more UV flux than TRAPPIST-1f leading to more oxygen radicals from CO, CO$_2$, H$_2$O, etc., photo-dissociation (see the strong decrease of the gas profiles at TOA). We can see in the TRAPPIST-1f/TRAPPIST-1e subplot that much more O$_2$ and O$_3$ are produced for TRAPPIST-1e above 20~km. Oxygen radicals consume the haze in TRAPPIST-1e, while for 1f, less hazes are  consumed and their concentration is larger. Underneath the TRAPPIST-1f thicker haze layer (from 85 km), O$_3$ and NO$_2$ are protected from photodissociation by the haze shielding. On the other side, TRAPPIST-1g is farther away and with less UV flux hence less hazes. Therefore, TRAPPIST-1f is at a sweet spot to maximize haze production which shields O$_3$ and NO$_2$ from photodissociation. This thicker haze layer for TRAPPIST-1f will also have a dramatic impact on the transmission spectra (see section \ref{sec:haze_spectra}).}

%FIGURE 8

\subsection{JWST simulated spectra: Impact of H$_2$O clouds and photochemical hazes \label{sec:haze_spectra}}
%FIGURE 9
\begin{figure}[ht!]
\centering
\resizebox{9cm}{!}{\includegraphics{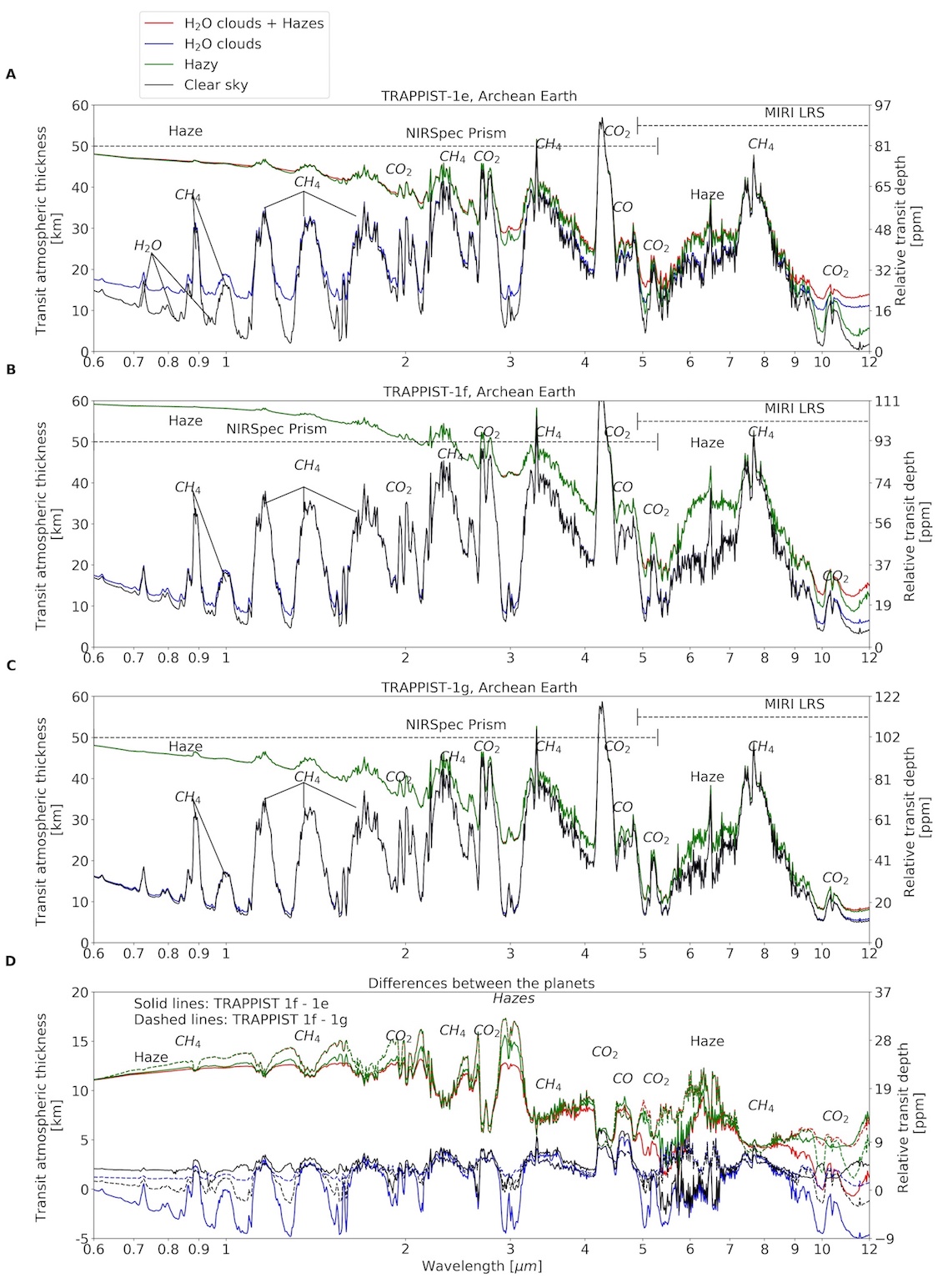}}
\caption{Simulated transmission spectra by JWST NIRSpec Prism and MIRI with R=300 for aquaplanets A) TRAPPIST-1e, B) 1f and C) 1g  with the Archean Earth atmosphere composition of \cite{Charnay2013} case B. Panel D) shows differences between planetary spectra.}
\label{fig:ArcheanSpectra}
\end{figure}

\begin{table*}
\centering
\caption{Relative transit depth (ppm), signal-to-noise ratio  for 1 transit (S/N-1) and number of transits to achieve a $5\ \sigma$ and $3\ \sigma$ detection for various spectral lines of the Archean Earth-like atmosphere of Charnay case B. Numbers in parentheses are for clear sky only while numbers without parentheses are the real values accounting for the impact of clouds and hazes. The hyphen represents the cases for which more than 100 integrated transits are needed \added{and the * mark denotes the values above the maximum number of transits per planet during JWST nominal lifetime mentioned in Table \ref{tab:TRAPPIST1}}.} \label{tab:transitArchean}
\resizebox{0.8\textwidth}{!}{\begin{minipage}{\textwidth}
\centering
\begin{tabular}{c c c  c }
\hline
\hline
 Planets & \added{TRAPPIST-1e} & \added{TRAPPIST-1f} & \added{TRAPPIST-1g}\\
\hline
Instrument & \multicolumn{3}{c}{\added{NIRSpec Prism (R=30)}}\\
\hline
Feature & \multicolumn{3}{c}{CH$_4$ $1.2\ \mu m$} \\
Depth [ppm] & 2(44) & 1(55) & 4(56)  \\
S/N-1  &  0.1(1.0) & 0.1(1.4) & 0.1(1.5)  \\
N transits  ($5\sigma$) & -(23) & -(13) & -(12)    \\
N transits  ($3\sigma$) & -(8) & -(5) & -(4)    \\
\hline
Feature  & \multicolumn{3}{c}{CO$_2$ $4.3\ \mu m$} \\
Depth [ppm] & 59(85) & 72(108) &   86(111)   \\
S/N-1  & 1.1(1.6)  & 1.4(2.2) & 1.7(2.3)  \\
N transits  ($5\sigma$) & 23(9) & 14(5) & 9(5) \\
N transits  ($3\sigma$) & 8(3) & 5(2) & 3(2) \\
\hline
Feature  & \multicolumn{3}{c}{CO $4.7\ \mu m$} \\
Depth [ppm] & 4(39) & 8(54) &   8(53)   \\
S/N-1  & 0.1(0.7)  & 0.1(1.0) & 0.1(1.0)  \\
N transits  ($5\sigma$) & -(59) & -(27) & -(25) \\
N transits  ($3\sigma$) & -(21) &-(10) & 76*(9) \\
\hline
Instrument &\multicolumn{3}{c}{\added{MIRI MRS (R=30)}}\\
\hline
Feature & \multicolumn{3}{c}{CH$_4$ $7.7\ \mu m$} \\
Depth [ppm] & 44(66) & 60(82) & 67(86)    \\
S/N-1  &  0.3(0.6) & 0.4(0.8) & 0.5(0.8)   \\
N transits  ($5\sigma$) & -(80) & -(44) & 98*(36)   \\
N transits  ($3\sigma$) & -(29) & 52(16) & 35(13)   \\
\hline
\hline
\end{tabular}
\end{minipage}}
\end{table*}

Figure \ref{fig:ArcheanSpectra} shows the TRAPPIST-1e, 1f and 1g transmission spectra, \added{and their relative differences,} for Charnay case B Archean Earth atmosphere with JWST NIRspec Prism and MIRI LRS. The hazes have a huge opacity down to the VIS/NIR which flattens most of the spectral features in the NIRSpec Prism range. \added{In the relative difference subplot, we can see that hazes are responsible for most of the differences between the spectra with TRAPPIST-1f spectrum being up to 15~km (or $\sim 30~ppm$) higher, as explanined in the previous section.}   \added{We can also see in Fig. \ref{fig:Reff} second subplot that the combination of the haze and H$_2$O ice clouds cover the whole atmospheric column for TRAPPIST-1f and therefore strongly absorbed the transmitted light.}\\
In the MIRI range, the hazes are clearly visible between 6 to $7\ \mu m$ \added{but at higher wavelengths their}  opacity progressively decreases (see Fig. \ref{fig:opticalprop}) and clouds become the largest source of opacity in the spectrum for TRAPPIST-1e and 1g (haze opacity in TRAPPIST-1f dominates the cloud opacity across the whole wavelength range).  \added{Similarly to Fig. \ref{fig:modernSpectra} clouds make a difference between the large cloud coverage of TRAPPIST-1e and the small cloud coverage of 1f and 1g.} 
The combined impact of clouds and hazes in the detectability of gaseous features is summarized in Table \ref{tab:transitArchean}. TRAPPIST-1g is the coldest and most distant of the three planets. \added{Less starlight heat the substellar point, muting the convection and therefore producing less clouds. Also farther away from the star, less UV photons are available to photodissociate CH$_4$ and form hazes. Therefore, TRAPPIST-1g} has the smallest amount of clouds and hazes allowing for less transits to detect the spectral lines than for the two other planets \added{but less transits are available during JWST lifetime (see Table \ref{tab:TRAPPIST1}).}  The most favorable band \added{to detect such atmosphere is}  CO$_2$ is at $4.3\ \mu m$ despite the presence of hazes at this wavelength, with only about 23, 14 and 9 transits required for a $5\ \sigma$ detection. The strength of the nearby CO feature is too weak to be detectable because of the continuum raised by hazes but also because, as mentioned previously, CO abundances may have been underestimated by fixing modern Earth mixing ratio and not predicting CO fluxes.  \added{CH$_4$ at 7.7 $\mu m$ is only detectable with MIRI at $3\ \sigma$ for TRAPPIST-1f and 1g (52 and 35 transits, respectively) while it will not be detectable at all with NIRSpec Prism  at 1.2 $\mu m$ because of the presence of hazes.} H$_2$O lines are either too shallow or are blended by CH$_4$ or CO$_2$ that they are undetectable. \\

\section{CO$_2$ atmospheres} \label{sec:CO2}
\subsection{Climate}
%FIGURE 10
\begin{figure}[ht!]
\centering
\resizebox{9cm}{!}{\includegraphics{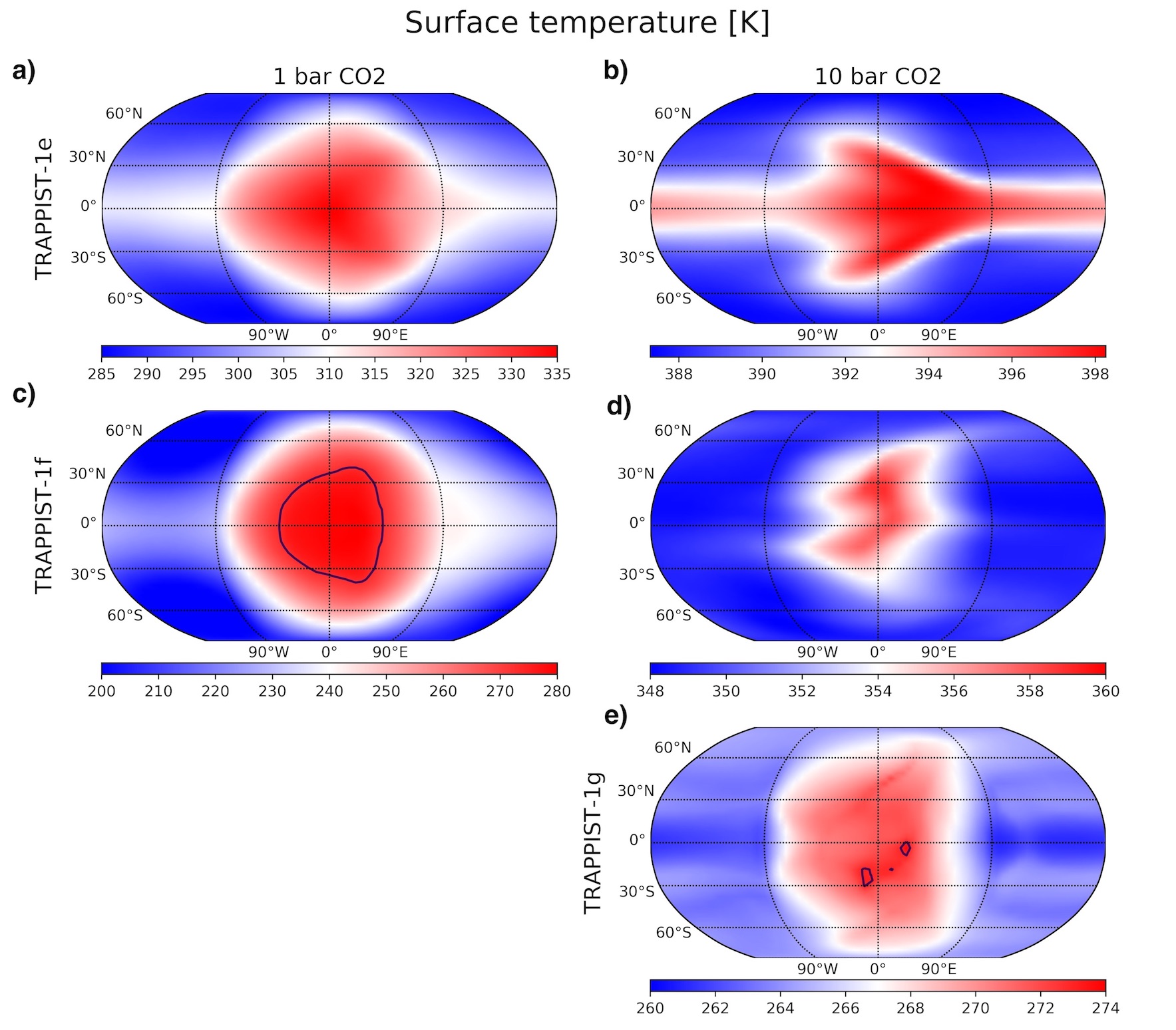}}
\caption{Surface temperature map in Kelvin (K) for aquaplanets TRAPPIST-1e (top row), TRAPPIST-1f (middle row) and TRAPPIST-1g (bottom row) at 1 (left column) and 10 (right column) bar of CO$_2$ surface pressure. TRAPPIST-1g at 1~bar is missing because the atmosphere would collapse on the night side. Note that the temperature scale is different for each of the subplot in order to highlight the so-called "lobster pattern" of TRAPPIST-1e at 10~bars which has a thermal amplitude of only 10~K. The blue line shows the sea-ice boundary. Note that TRAPPIST-1e at 1 and 10~bars are completely ice-free.}
\label{fig:CO2surftemp}
\end{figure}

Among the four rocky planets of our solar system, CO$_2$ is the dominant gas on two of them (Venus and Mars), and is thought to have been a dominant gas in early Earth's atmosphere, in particular during the Hadean epoch \citep{Zahnle2010}. Therefore, it is reasonable to think that CO$_2$ atmospheres may be common in other planetary systems as well. \cite{deWit2018,Moran2018} have shown that if the TRAPPIST-1 planets have an atmosphere, they should be free of  low mean-molecular weight gases such as hydrogen or helium in absence of haze. This raises a possibility of high mean molecular weight species, such as CO$_2$, as a possible constituent.
For each planet in the habitable zone of TRAPPIST-1 (i.e. planet~e, f and g), we used LMD-G to simulate CO$_2$-dominated atmospheres with 1 and 10 bar surface pressures. However, we were not able to successfully simulate the 1 bar CO$_2$ atmosphere for TRAPPIST-g, because the atmospheric temperature on the night side is cold enough that CO$_2$ condenses \added{(below 194~K at 1~bar)} on the surface, resulting in atmospheric collapse. TRAPPIST-1g retaining 1 bar or less of CO$_2$ is therefore highly unstable and unlikely to \added{occur as also found in}  \cite{Turbet2018}. Figure \ref{fig:CO2surftemp} shows the surface temperature maps, averaged over 10 orbits, of 1 bar (left column) and 10 bars (right column) CO$_2$-dominated atmospheres for TRAPPIST-1e (top row), f (middle row) and g (bottom row). Surface temperatures and integrated columns of condensed species are reported in Table~\ref{tab:climateCO2}. As for the modern Earth-like atmosphere, mean surface temperatures predicted by the GCM for the 10 bars cases agree with other GCM simulations \citep{Wolf2017} but are much higher than the one predicted with 1-D climate model of \cite{Lincowski2018} for their 10 bars Venus -like atmospheres primarily due to the cooling of the highly reflective sulfuric acid aerosols (not included in \cite{Wolf2017} nor in our study).
Note that none of the 10~bars simulations are cold enough to have CO$_2$ condensation at the surface (below 233.6~K), in agreement with \cite{Turbet2018}. At 1 bar, we can see that TRAPPIST-1e is ice-free, while TRAPPIST-1f is an "eye-ball" planet, with an open ocean restricted to the substellar region, roughly between -40 and +40 longitude East and -40 and +40 latitude North. In both cases, the surface temperature contrast between the substellar and anti-substellar region is roughly 100~K. At 10 bar surface pressure, the atmosphere is very efficient to transport the heat and the contrast is only on the order of 10~K for TRAPPIST-1e, f and g. Very interestingly, because of the faster rotation period of TRAPPIST-1e (6.1 days) a so-called "lobster pattern" appears, which is usually seen when a dynamic ocean is coupled to the atmosphere \citep{Hu2014,DelGenio2019}. This asymmetric pattern of surface temperature is due the combination of a Rossby wave, West to the substellar point, moving the warm air away from the equator, and a Kelvin wave, East of the substellar point, progressing exclusively in the longitude-altitude plane. While ocean heat transport is not included in our simulation, the combination of the dense atmosphere and fast rotation rate are responsible for this pattern.\\

\begin{table*}
\caption{Surface temperatures (TS) and integrated column of condensed species for the CO$_2$ dominated atmospheres. Values with an asterisk are averaged around the terminator only.} \label{tab:climateCO2}
\resizebox{0.9\textwidth}{!}{\begin{minipage}{\textwidth}
\centering
\begin{tabular}{c c c  c c  c }
\hline
\hline
%\added{Parameters} & \multicolumn{5}{c}{Planets}\\
 Planet & \multicolumn{2}{c}{\added{TRAPPIST-1e}} & \multicolumn{2}{c}{\added{TRAPPIST-1f}} & \added{TRAPPIST-1g}\\
 & \multicolumn{2}{c}{CO$_2$ dominated} & \multicolumn{2}{c}{CO$_2$ dominated} & CO$_2$ dominated\\
Pressure & 1 bar & 10 bar & 1 bar & 10 bar  &  10 bar \\
TS mean [$K$] & 303 & 392  & 230 & 350  & 266 \\
TS min [$K$] & 285 & 387 &  194 & 348  & 261  \\
TS max [$K$]  &  335 & 398 & 281 & 359  & 274 \\
H$_2$O liq* [$10^{-3}kg\cdot m^{-2}$] & 61.3 & 28.3 & $1.0\times 10^{-2}$ & 26.7 & $2.0\times 10^{-1}$ \\
H$_2$O ice* [$10^{-3}kg\cdot m^{-2}$] & 9.9 & 12.4 & 4.3 & 9.4 & 5.7\\
CO$_2$ ice* [$10^{-3}kg\cdot m^{-2}$] & 0.0 & 0.0 & $3.1\times10^{-2}$ & $6.6\times10^{-1}$ & 90.0 \\
\hline
\hline
\end{tabular}
\end{minipage}}
\end{table*}

\begin{figure}[h!]
\centering
\resizebox{9cm}{!}{\includegraphics{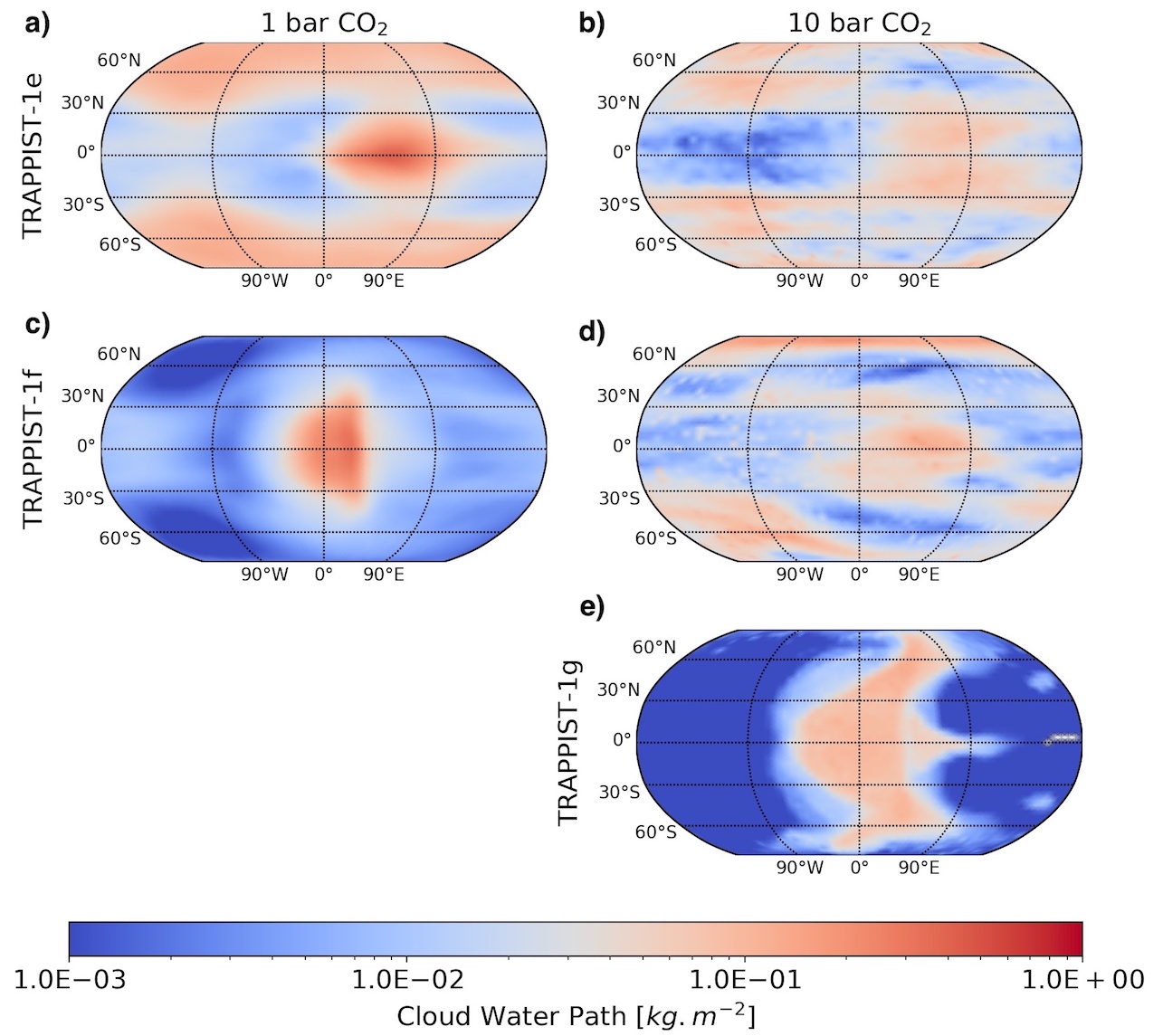}}
\caption{Integrated H$_2$O column in $kg.m^{-2}$ for aquaplanets TRAPPIST-1e (top row), TRAPPIST-1f (middle row) and TRAPPIST-1g (bottom row) at 1 (left column) and 10 (right column) bar of CO$_2$ surface pressure. TRAPPIST-1g at 1~bar is missing because the atmosphere has condensed to the night side leading to the crash of the simulation.}
\label{fig:CO2_H2Ocol}
\end{figure}

Figure \ref{fig:CO2_H2Ocol}  shows the integrated columns of H$_2$O condensates (liquid and ice). The largest cloud coverage was recorded for both TRAPPIST-1e and -1f at 1~bar, with a large cloud deck due to the strong convection and shifted eastward of the substellar point \citep{Kopparapu2017,Yang2014} for TRAPPIST-1e due to the fast rotation. Note that for TRAPPIST-1f, the rotation is slower and the cloud deck is more centered toward the ice-free substellar region.\\
At 10~bar surface pressure, TRAPPIST-1e and -1f are so warm that the huge amount of water vapor brought to the atmosphere leads both inefficient radiative cooling and strong solar absorption in the low atmosphere, causing a net radiative heating of the layers near the surface; subsequently, this radiative heating creates a strong temperature inversion encompassing the entire planet, stabilizing the low atmosphere against convection, including at the substellar point \citep{Wolf2015}. Indeed, inversion layers are intrinsically stable against vertical mixing; without a deep convection carrying moisture up from the boundary layer, no substellar cloud deck is formed, and instead, the skies are relatively clear despite the enormous amount of water vapor in the atmosphere.  \added{From Fig. \ref{fig:Reff} last subplot we can see that TRAPPIST-1f atmosphere forms H$_2$O liquid clouds from 0 to about 10~km and H$_2$O ice clouds from 10 to 60~km. Very thin CO$_2$ clouds also formed at high altitudes but are not shown in the figure.} TRAPPIST-1g is much colder, almost fully ice-covered except at a few spots near the substellar region (see Fig. \ref{fig:CO2surftemp} bottom row) where some water can evaporate from the ocean and form relatively thin clouds with a H$_2$O water cloud column of about 0.1 $kg\cdot m^{-2}$.

%FIGURE 11

\begin{figure}[!ht]
\centering
\resizebox{9cm}{!}{\includegraphics{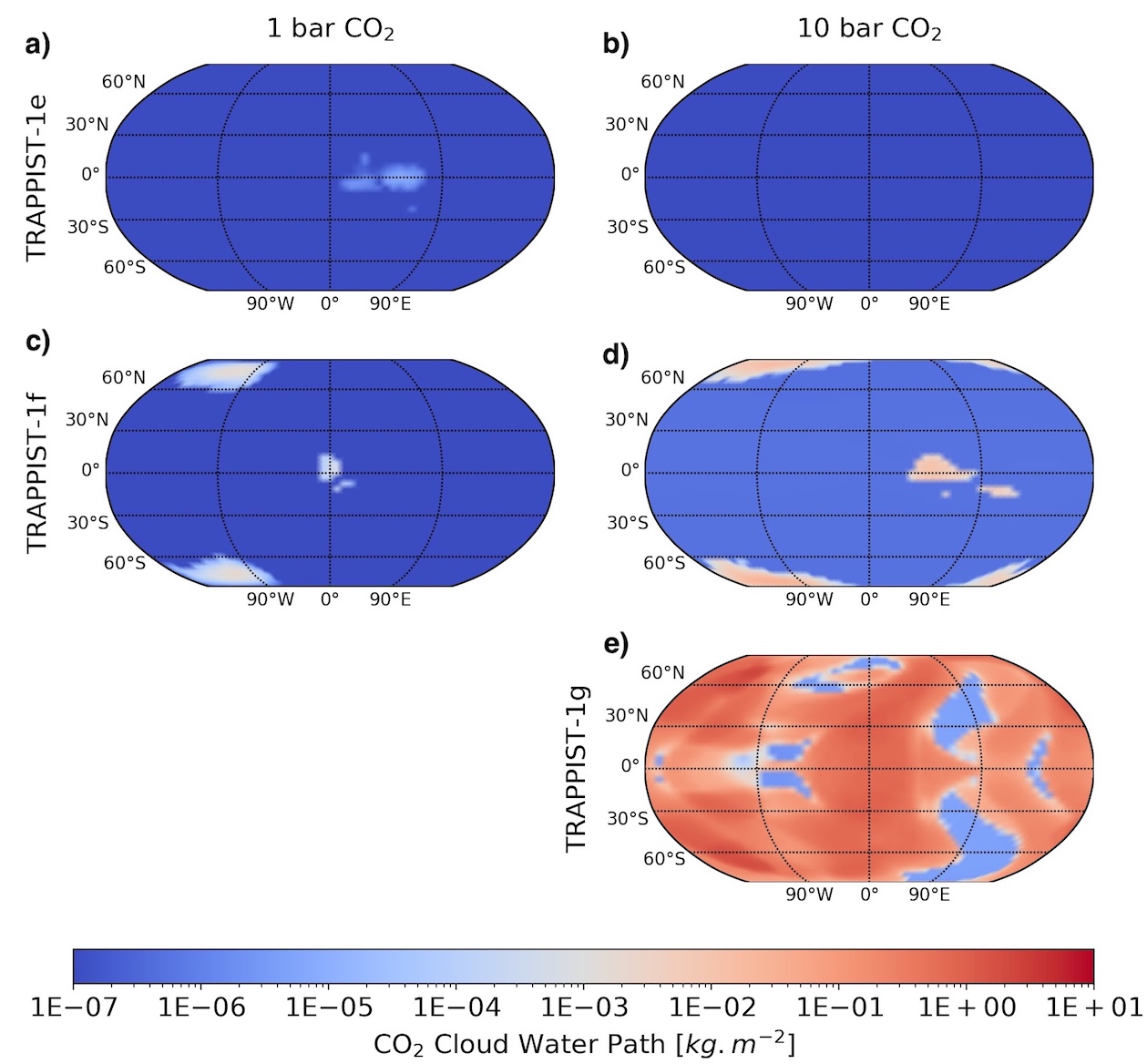}}
\caption{Integrated CO$_2$ column in $kg.m^{-2}$ for aquaplanets TRAPPIST-1e (top row), TRAPPIST-1f (middle row) and TRAPPIST-1g (bottom row) at 1 (left column) and 10 (right column) bar of CO$_2$ surface pressure. TRAPPIST-1g at 1~bar is missing because the atmosphere has condensed to the night side leading to the crash of the simulation. All planets are aquaplanets.}
\label{fig:CO2_CO2col}
\end{figure}

Figure \ref{fig:CO2_CO2col} shows the integrated column of CO$_2$ ice. TRAPPIST-1e is too warm at 1 and 10 bars to have significant CO$_2$ condensation in the atmosphere. For TRAPPIST-1f, CO$_2$ starts to condense in two cold traps \citep{Leconte2013b} at 1 bar at a symmetric position around longitude $-120^\circ$ and latitudes $\pm 80^\circ$
 and between 30 and 50~km (see Fig.\ref{fig:Reff}) but their position can slightly vary, due to planetary-scale equatorial Kelvin and Rossby wave interactions \citep{Showman2011}. Also, we notice that for a thicker atmosphere (10~bar) these two colds traps tend to move westward and toward highest latitudes and two others are at longitude $+150^\circ$ and latitudes $\pm 90^\circ$. Note that a few spots of CO$_2$ condensate appear eastward of (TRAPPIST-1e) and at (TRAPPIST-1f) the substellar point. This is due to a local temperature minimum at p=67 mbar near the substellar point marking the top of the ascending circulation branch \citep{Carone2014,Carone2015,Carone2018}. These CO$_2$ clouds near the substellar point would likely disappear due to the shortwave absorption of the CO$_2$ ice crystals but the radiative effect of CO$_2$ is not taken into account in our simulations.\\

%FIGURE 12

\subsection{JWST simulated spectra: Impact of H$_2$O and CO$_2$ clouds}

\begin{figure}[h!]
\centering
\resizebox{9cm}{!}{\includegraphics{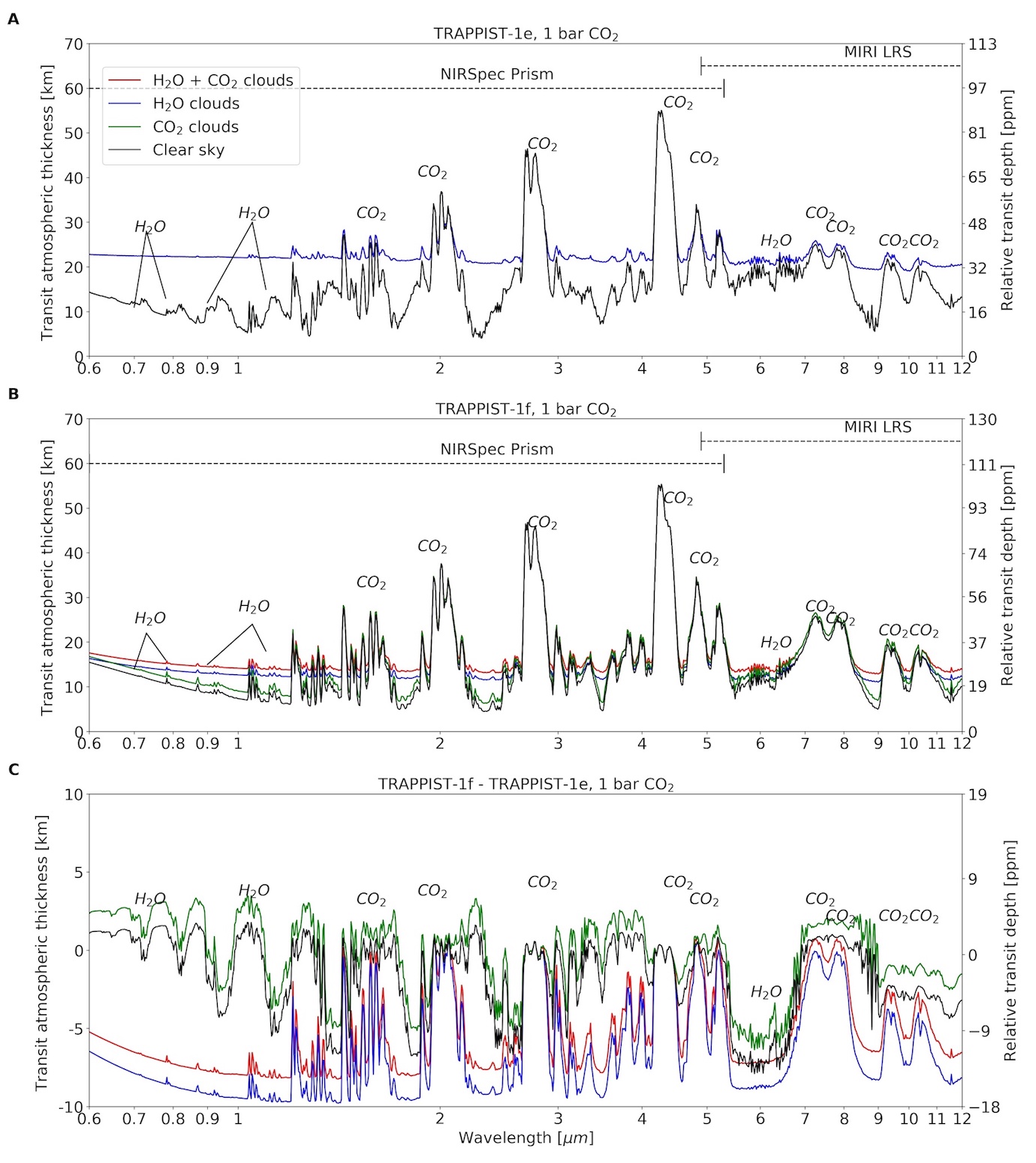}}
\caption{Simulated transmission spectra by JWST NIRSpec Prism and MIRI with R=300 for aquaplanets A) TRAPPIST-1e  and B) 1f  with 1 bar of surface pressure of CO$_2$. Panel D) shows differences between planetary spectra.}
\label{fig:1barCO2_NIRSpec}
\end{figure}

\begin{figure}[ht!]
\centering
\resizebox{9cm}{!}{\includegraphics{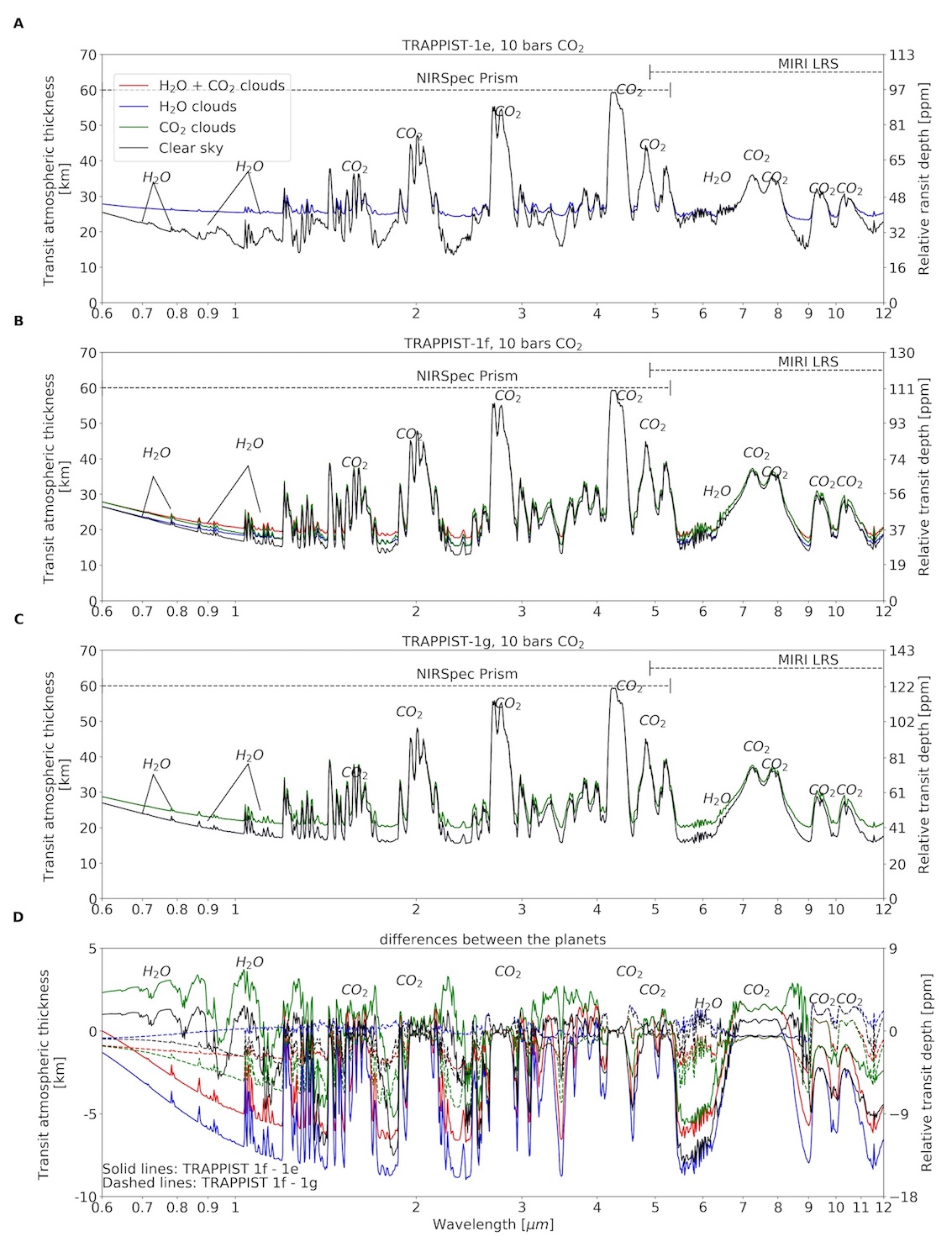}}
\caption{Simulated transmission spectra by JWST NIRSpec Prism and MIRI with R=300 for aquaplanets A) TRAPPIST-1e, B)  1f and C) 1g  with 10~bars of surface pressure of CO$_2$. Panel D) shows differences between planetary spectra.}
\label{fig:10barCO2_NIRSpec}
\end{figure}

\begin{figure}[ht!]
\centering
\resizebox{9cm}{!}{\includegraphics{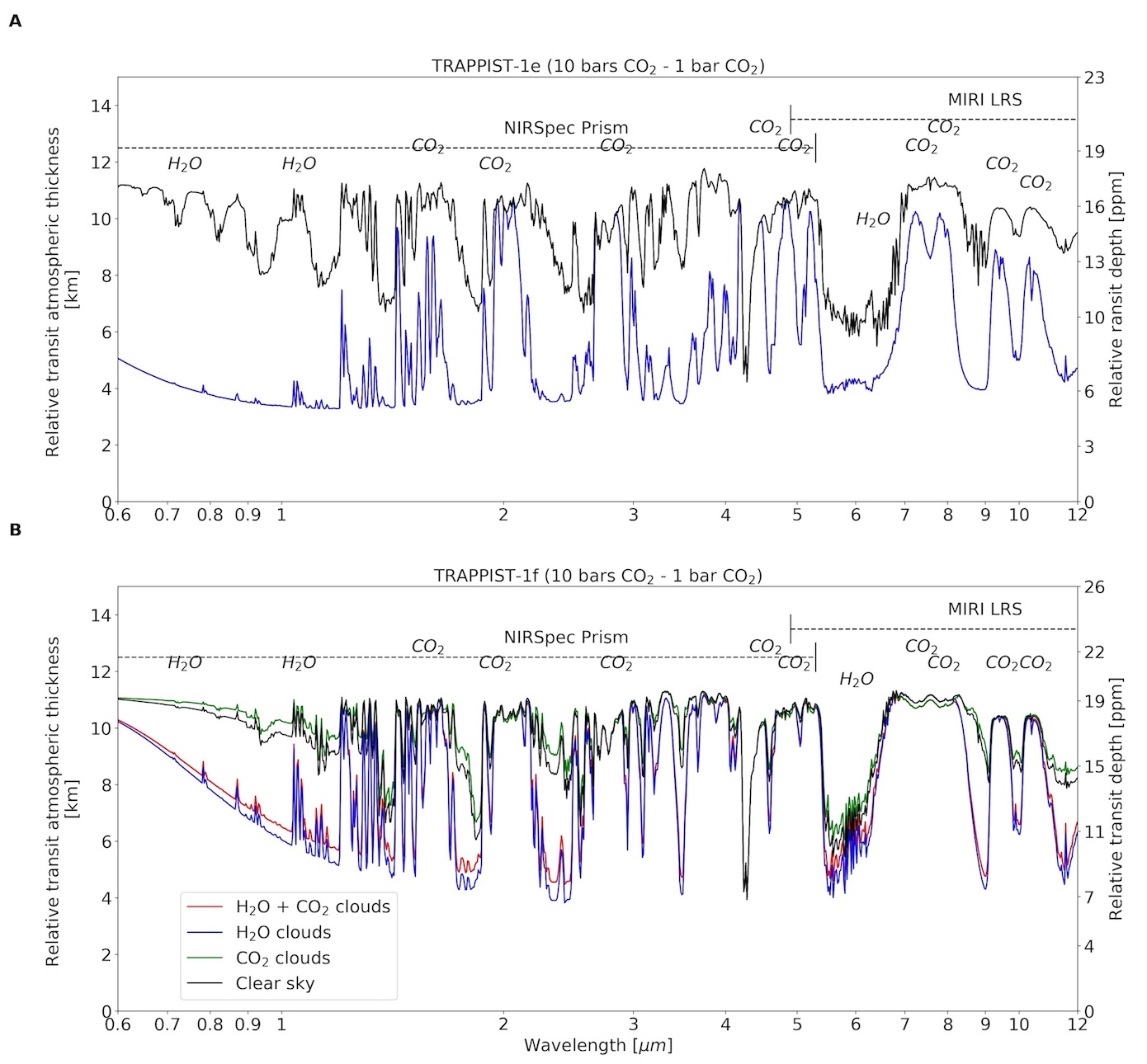}}
\caption{Difference between the spectra of the 10~bars and 1~bar CO$_2$ surface pressure for aquaplanets A) TRAPPIST-1e  and B) 1f.}
\label{fig:10-1barCO2_NIRSpec}
\end{figure}

Figures \ref{fig:1barCO2_NIRSpec} and \ref{fig:10barCO2_NIRSpec} show  JWST NIRSpec Prism and MIRI simulated transmission spectra for TRAPPIST-1e, 1f and -1g at 1 and 10 bar CO$_2$ surface pressures, respectively. In addition, \added{the relative difference between the transmission spectra for the 10 and 1 bar surface pressure atmospheres is shown in Fig. \ref{fig:10-1barCO2_NIRSpec} for TRAPPIST-1e and 1f.} The relative transit depth, the signal-to-noise ratio (S/N) for 10 transits and the number of transits for a $5\ \sigma$ and $3\ \sigma$ detection are reported in Table \ref{tab:transitCO2}.\\
%FIGURE 13

%FIGURE 14

First, we can see \added{in Fig. \ref{fig:1barCO2_NIRSpec}} that water clouds  produce a considerable flattening of the spectra of TRAPPIST-1e, suppressing H$_2$O lines and leading to a continuum level at about 22~km in the 1~bar case. Around the terminator, the average liquid and ice water contents (LWC and IWC, respectively) are equal to $4.1\times10^{-6}$ and $6.5\times10^{-7}$ $kg\cdot m^{-3}$ for TRAPPIST-1e and 1f, respectively (see Table \ref{tab:climateCO2}). \added{The CO$_2$ clouds slightly raise the continuum in TRAPPIST-1f. The differences between the two spectra in the clear sky atmosphere is due to the stronger H$_2$O lines in 1e, while the CO$_2$ lines are very similar. In the cloudy atmosphere, TRAPPIST-1e have the higher continuum and therefore the smaller absorption lines due to H$_2$O clouds.} \\ 

\added{Figure \ref{fig:10barCO2_NIRSpec} shows a transition between  TRAPPIST-1e to 1f and 1g where the TRAPPIST-1 flux received by the planets is reduced. Lower fluxes imply colder surface temperatures leading to less water evaporation, lower water mixing ratio in the atmosphere and shallower water lines. When less water vapor exists,  less water clouds are produced, reducing the flattening of water lines and therefore paradoxically improving their detection. Indeed, we can see in Table \ref{tab:transitCO2} that the differences in transit depth between the clear sky and cloudy values is the smallest for TRAPPIST-1g. TRAPPIST-1e is too warm to have CO$_2$ cloud to condense (see also Fig. \ref{fig:CO2_CO2col}) but has an opaque H$_2$O cloud deck at about 28~km. TRAPPIST-1f has much less H$_2$O clouds and a few CO$_2$ clouds slightly raise the continuum. Finally, the spectrum of the colder TRAPPIST-1g is the most impacted by CO$_2$ clouds while the few H$_2$O clouds are below the atmospheric refraction limit ($\sim 15~km$) and are therefore not observable in the spectrum. In the relative difference subplot, we can see that the continuum and the H$_2$O lines are the major differences between the spectra of the three planets, while the intensity of the CO$_2$ lines are fairly similar.}\\

\begin{figure}[h!]
\centering
\resizebox{9cm}{!}{\includegraphics{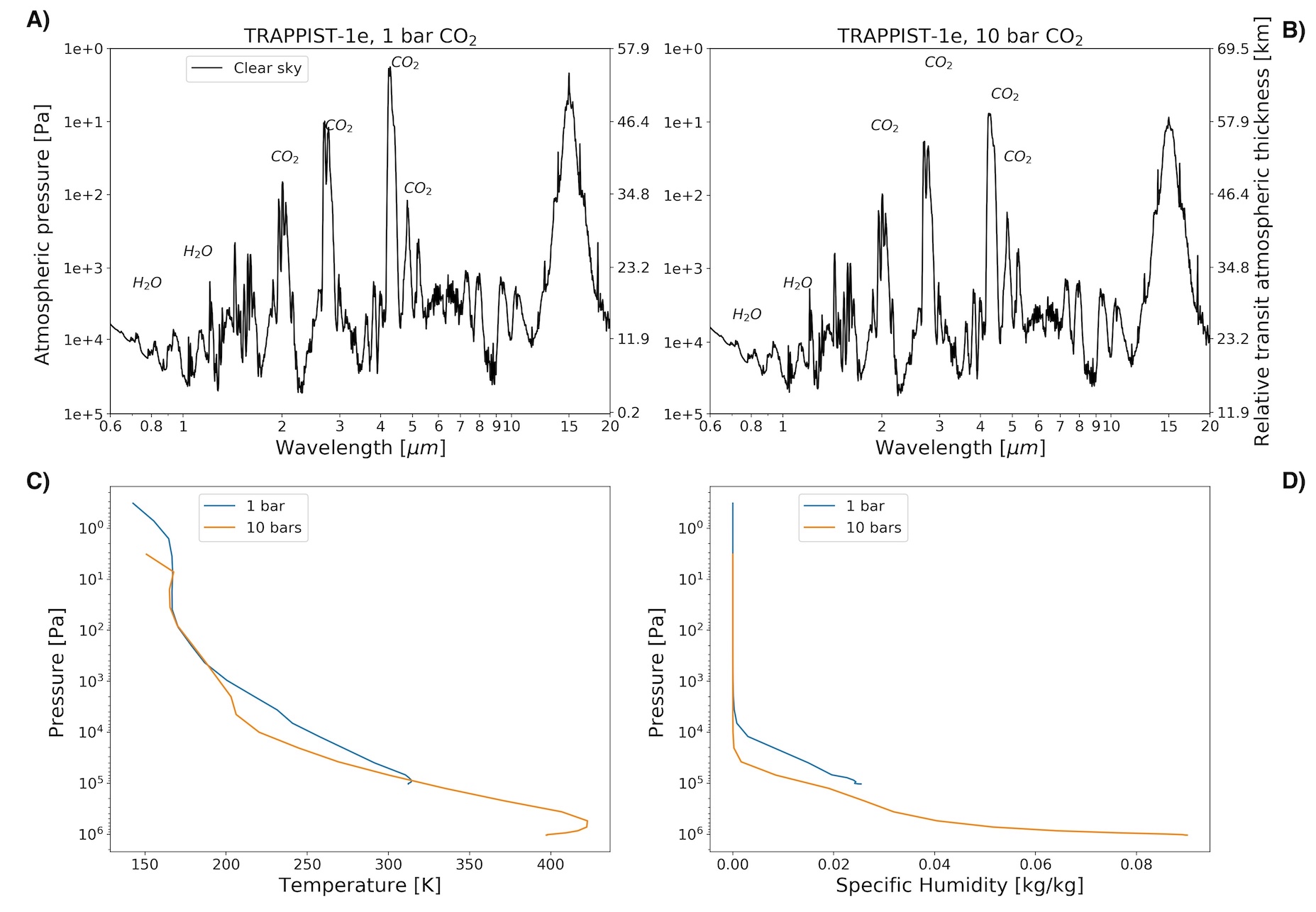}}
\caption{Simulated transmission spectra by JWST NIRSpec Prism and MIRI with R=300 for the aquaplanet TRAPPIST-1e with  A) 1~bar and B) 10~bars of CO$_2$ surface pressure. C) Atmospheric temperature and D) specific humidity, averaged at the terminator, as a function of the atmospheric pressure for 1 and 10 bars CO$_2$ surface pressures.}
\label{fig:1_10barCO2_NIRSpec}
\end{figure}

 At 10~bar surface pressure, the mean surface temperature of TRAPPIST-1e is very high, 392~K, \added{compared to 303~K at 1~bar}. At this temperature, the continuum of water vapor in the low atmosphere is opaque to the infrared radiation, and the continuum is pushed toward higher altitudes, even in clear sky, up to about 15~km. Figure \ref{fig:10barCO2_NIRSpec} shows that this results in a reduction of the relative transit depth of every gas including CO$_2$ \added{compared with the spectra at 1~bar surface pressure (Fig. \ref{fig:1barCO2_NIRSpec}).
 This can also be seen in Fig. \ref{fig:1_10barCO2_NIRSpec}, in unit of atmospheric pressure, where clear sky absorption lines are  shallower at 10~bar than at 1~bar surface pressure. This is counter-intuitive because larger gas pressures are expected to produce stronger absorption lines, as long as the temperatures are assumed to be constant. However, as we can see in the bottom left plot of Fig. \ref{fig:1_10barCO2_NIRSpec}, the  temperature at the atmospheric pressures ($10^5-10^3$ Pa) where the lines are emitted (where the atmosphere goes from optically thick to optically thin) is colder at 10~bars than at 1~bar. Indeed, at 10~bars, the specific humidity of the atmosphere is much lower (bottom right plot) and the  temperature lapse rate is getting steeper (closer to the dry adiabat, \cite{Wolf2015}) leading to a faster decrease of temperature. Yet, pressure broadening half-width is inversely proportional to temperature. So at the colder emission temperatures of the 10~bar atmospheres, the line broadens and the peak intensity, relative to the continuum, becomes lower. However, because at 10~bar surface pressure the whole spectrum is raised toward higher altitudes, it shows larger transit atmospheric thickness and transit depth relative to the ground than at 1~bar surface pressure. As a result, Fig. \ref{fig:10-1barCO2_NIRSpec} shows the larger \deleted{relative} transit atmospheric thickness and transit depth at 10~bars in clear sky (more than 10~km or 15~ppm)  for TRAPPIST-1e and 1f because of the continuum IR opacity while the cloudy (H$_2$O) spectra shows only few kilometers of differences because clouds form roughly at the same altitude. Note that because with real observations we will not be able to differentiate a ground level from a cloud deck in this wavelength range, the overall result will be a reduction of the relative intensity of the absorption lines for the warmer 10~bar CO$_2$ surface pressure case.} \\

 Similarly to the atmospheres with modern Earth and Archean Earth boundary conditions, we can see that H$_2$O lines are not detectable  at $3\ \sigma$ or $5\ \sigma$ in less than 100 transits for the cloudy scenario. MIRI does not performed better, with no detectable H$_2$O lines. On the contrary, the well-mixed CO$_2$ is barely affected  by the presence of clouds in the line region, because enough of it remains above the cloud deck. Because the continuum level is raised by the presence of clouds, the transit atmospheric thickness and transit depth of CO$_2$ is also reduced.  CO$_2$ at 4.3 $\ \mu m$ has a transit depth of the order of 50 to 80~ppm and could be detected with NIRSpec  from 10 to 30 transits at $5\ \sigma$ confidence level. Note that the number of transit at $5\ \sigma$ for the 10 bar clear sky case  of TRAPPIST-1e (22) is the same as the one estimated by \cite{Lustig_Yaeger2019} for NIRSpec Prism sub512 mode.\\

%FIGURE 15

%FIGURE 16

\begin{table*}
\centering
\caption{Relative transit depth (ppm), signal-to-noise ratio  for 1 transit (S/N-1) and number of transits to achieve a $5\ \sigma$ and $3\ \sigma$ detection for various spectral lines of the CO$_2$ -dominated atmosphere. Numbers in parentheses are for clear sky only while numbers without parentheses are the real values accounting for the impact of clouds. The hyphen represents the cases for which more than 100 integrated transits are needed and the * mark denotes the values above the maximum number of transits observable per planet during JWST nominal lifetime mentioned in Table \ref{tab:TRAPPIST1}.} \label{tab:transitCO2}
\resizebox{0.8\textwidth}{!}{\begin{minipage}{\textwidth}
\centering
\begin{tabular}{c c c  c c  c }
\hline
\hline
 Planets & \multicolumn{2}{c}{\added{TRAPPIST-1e}} & \multicolumn{2}{c}{\added{TRAPPIST-1f}} & \added{TRAPPIST-1g}\\
Pressures & 1 bar & 10 bar & 1 bar & 10 bar  &  10 bar \\
\hline
%Instrument & \multicolumn{5}{c}{\added{NIRSpec Prism (R=30)}}\\
%\hline
Feature & \multicolumn{5}{c}{H$_2$O $1.4\ \mu m$} \\
Depth [ppm] & 3(18) & 3(14) & 4(12)   & 5(9)  &  5(10) \\
S/N-1  &  0.1(0.4) & 0.1(0.3) & 0.1(0.3)  & 0.1(0.2)  & 0.1(0.2) \\
N transits  ($5\sigma$) & -(-) & -(-) & -(-)  &  -(-)  & -(-) \\
N transits  ($3\sigma$) & -(54) & -(82) & -(-)  &  -(82*)  & -(-) \\
\hline
Feature  & \multicolumn{5}{c}{CO$_2$ $4.3\ \mu m$} \\
Depth [ppm] & 61(86) & 50(56) & 80(96)  & 60(66)  & 67(74)  \\
S/N-1  & 1.2(1.6)  & 0.9(1.1) & 1.6(1.9) & 1.2(1.3)  & 1.4(1.6) \\
N transits  ($5\sigma$) & 19(9) & 28(22) &10(7) & 17(15) & 13(10) \\
N transits  ($3\sigma$) & 7(3) & 10(8) & 4(2) & 6(5) & 5(4) \\

\hline
\hline
\end{tabular}
\end{minipage}}
\end{table*}

\section{Discussion}\label{sec:discussion}
\subsection{Noise and detectability}
In this section, we discuss the different noise sources that can impact JWST observations and the detectability of an atmosphere and/or of any gaseous feature.\\
The large aperture of JWST (6.5~m) will allow us to quickly acquire a significant number of photons after a few transits while the noise from the source (N$_{source}$) will largely dominate the total noise (N$_{total}$). \added{In a photon-limited noise scenario, the noise can be represented by a "white noise", which decreases when acquiring more photons. For transmission spectroscopy, it can be expressed by $1 / \sqrt{X}$ or X$^{-0.5}$ with X the number of transit.  However,} every instrument suffers from a background red noise (of low frequency) which is a measurement error in addition to the frequency coming from the white noise (photon, reading, dark, etc). This red noise comes mainly from the systematic effects that affect the measurements, e.g., the fact that the pixels are not perfectly homogeneous (intra-pixel gain variability \citep{Knutson2008,Anderson2011} and that the telescope does not track perfectly, resulting in a position-dependent low frequency noise which can be modeled but \added{will lead to greater uncertainty} (because the model is never a perfect representation of the noise, and its parameters have their \added{own errors that vary in magnitude with the measurement itself). A red noise is expected to stay constant or decrease very slowly with the number of transits, and can be represented by a small X exponent. An intermediate scenario of a "pink noise", as often used to describe sounds, is when the noise also decreases with X but slower than for a white noise. This is the realistic scenario considered here.}\\

According to \cite{Greene2016},  instrumental noise (introduced by decorrelation residuals) produces systematic noise floors that do not decrease when acquiring more photons (with a larger aperture and/or more integration time), \added{like a red noise}. In HST WFC3 observations of  GJ 1214 \cite{Kreidberg2014}, the errors obtained from integrated 15 transits are however in perfect agreement with a modeled "pure white noise", indicating a low noise (30~ppm) and decay close to \added{X$^{-0.5}$}. \cite{Tsiaras2016} report the most precise transmission spectrum for a planet (55 Cancri e) with a single visit with HST WFC3 reaching 20-30~ppm precision over 25 channels. In the infrared with the Spitzer space telescope, values as low as 65~ppm have been achieved \citep{Knutson2009}. If we observe a large number of transits,  the difference in frequency between the systematic effects and the orbit of the planet will approach the reduction in \added{X$^{-0.5}$} of the white noise, but without ever reaching it \added{(pink noise)}. To suppose a fixed background (red) noise as in \cite{Greene2016} implies to neglect this decrease.
However, the fact that a noise floor better than 30~ppm has not been achieved yet is not due to the precision limit of instruments like HST WFC3, but instead to the fact that no one has ever accumulated enough high S/N transits. Yet, it is only by accumulating a large number of transits of the same object with JWST that we will know if the instruments can do better, and measure the value of their background noise and the profile of its decay as a function of the number of transits. Note that this decay is very poorly characterized in IR spectrophotometry because to quantify it requires high S/N observations and many transits observed.  \\
Another way to estimate what we can expect to achieve as estimated precision with JWST is to look at the accuracy reached by Spitzer or WFC3 at very high S/N, in photometry rather than in spectrophotometry. For HD 219134, \cite{Gillon2017b} have obtained a 20~ppm  noise  with only 2 transits at $4.5\ \mu m$, with a much less homogeneous InSb detector than the NIRSpec or WFC3 HgCdTe detector. Compared to the expected white noise, this produces a red noise of less than 10~ppm, despite systematics of about 1000~ppm amplitude. \\

\begin{figure}[h!]
\centering
\resizebox{9cm}{!}{\includegraphics{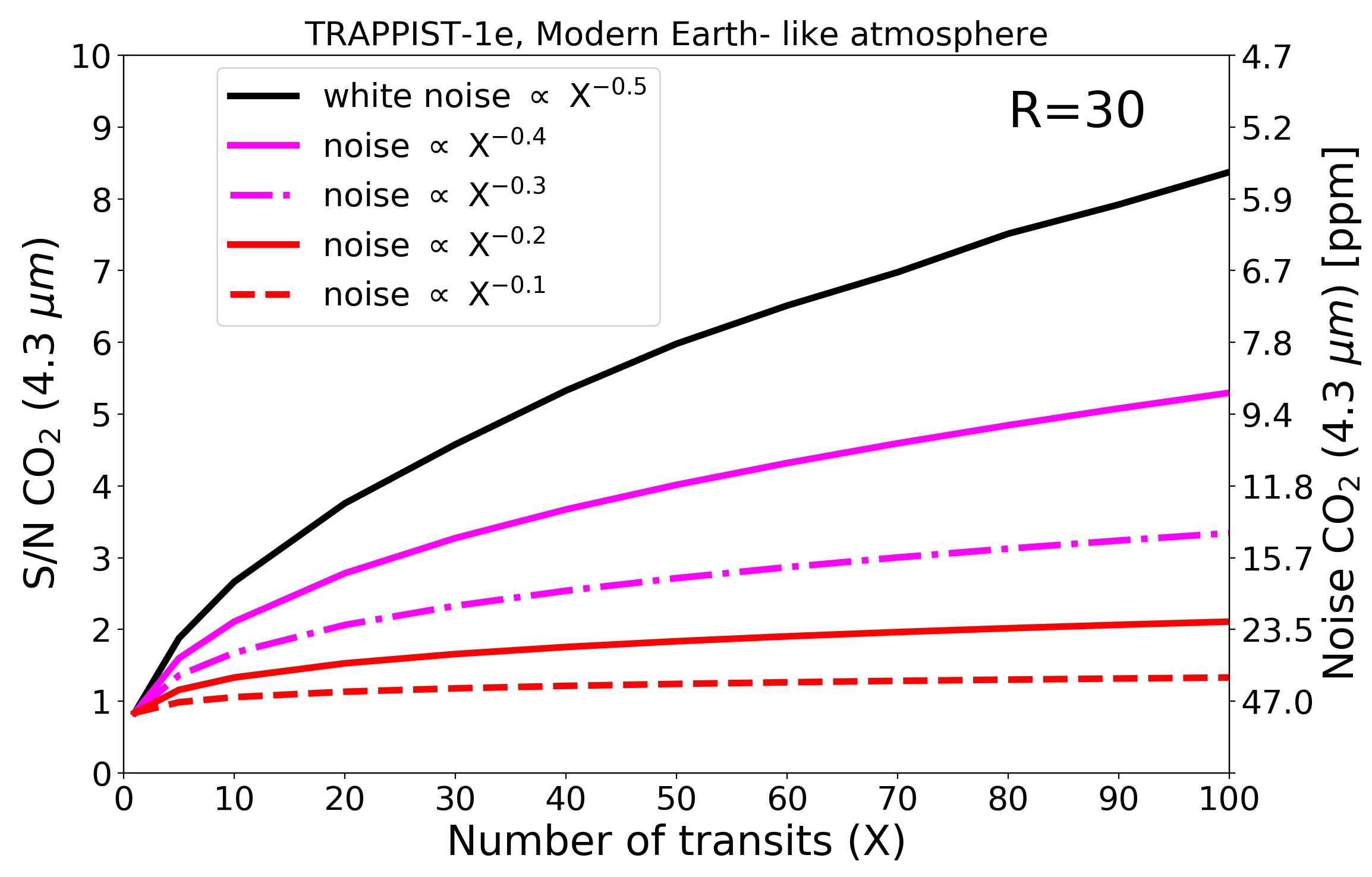}}
\caption{Signal-to-noise ration (S/N) (left y axis) and noise (right y axis) for the CO$_2$ line at 4.3 $\mu m$ of the modern Earth-like atmosphere simulated for JWST NIRSpec prism at R=30 as a function of the number of transits. \added{White noise ($X^{-0.5}$ with X the number of transit) is represented by the black curve, while S/N with smaller X exponents are represented in pink (smaller increase of S/N with X) and red (almost constant with X) colors.}}
\label{fig:noise}
\end{figure}

 Figure \ref{fig:noise} shows the S/N (left Y-axis) and noise (right Y-axis) for the CO$_2$ line at 4.3 $\mu m$ of the modern Earth-like simulation as a function of the number of transits. \added{We can see that when a white noise is assumed (black curve) a S/N of 5 ($5\ \sigma$ detection) is reached in about 35 transits (see also Table \ref{tab:transitmodern}). If the noise decreases slower (-0.5 $<$ X exponent $<$ 0.0) than for a white noise, $5\ \sigma$ detection will require more transits. We show here that the noise exponent should not be greater than -0.4 to reach  $5\ \sigma$ detection in less than 100 transits.}\\

%FIGURE 17

\added{\cite{Deming2009, Greene2016} have assumed 1 $\sigma$  noise floors for NIRSpec prism ($\lambda= 0.6 - 5\ \mu m$) and  MIRI LRS ($\lambda= 5.0 - 11\ \mu m$) of 20 and 50~ppm, respectively. A 20~ppm noise floor for NIRSpec would correspond to the solid red line in Fig. \ref{fig:noise} of a noise depending on X$^{-0.2}$. However, we consider these values to be conservative. Indeed, unlike HST or Spitzer,  the detector systematic behaviour for exoplanet spectroscopy can be studied prior to the launch for JWST. This advantage, combined with the continuing improvement of data reduction techniques, should minimize the detector systematics for JWST. Following the various arguments explained above, we consider half of the \cite{Deming2009, Greene2016} noise floor values, i.e. 10 and 25~ppm $1\ \sigma$ optimistic  noise floors for NIRSpec prism and MIRI, respectively.}
Meanwhile, in this study we consider the significance of a detection of an atmosphere (whatever the gas) at a $3\ \sigma$ confidence level  but the detection of a specific biosignature gas such as O$_2$, O$_3$, CH$_4$ or even H$_2$O at $5\ \sigma$. The a priori noise floors should therefore be scaled accordingly by the factor of the confidence level. Table \ref{tab:floor} shows the various noise floors as a function of the significant level considering either 20 and 50~ppm or 10 and 25~ppm, for NIRSpec and MIRI, respectively. \\

\begin{table}
\centering
\caption{A priori noise floors as a function of the detection confidence level from the conservative and optimistic $1\ \sigma$ noise floor estimations (first and second column, respectively, for each confidence level).} \label{tab:floor}
\begin{tabular}{c c c|  c c| c c }
\hline
\hline
  & \multicolumn{6}{c}{\added{Noise floors}}\\
& \multicolumn{2}{c}{$1\ \sigma$} & \multicolumn{2}{c}{$3\ \sigma$} & \multicolumn{2}{c}{$5\ \sigma$} \\
 NIRSpec prism & 10 & 20 & 30 & 60 & 50 & 100 \\
 MIRI LRS & 25 & 50 & 75 & 150 & 125 & 250 \\
\hline
\hline
\end{tabular}
\end{table}

 Table  \ref{tab:transitmodern} (modern Earth atmosphere), Table \ref{tab:transitArchean} (Archean Earth atmosphere) and Table \ref{tab:transitCO2} (CO$_2$ rich atmospheres) \added{show the number of transits needed to detect various gas features assuming a white noise and therefore without considering a noise floor. However, only the relative transit depth of the lines larger than the noise floor could be detected. For each gas feature allowing a possible 3 or $5\ \sigma$ detection,  the relative transit depth should be compared with the estimated noise floor values of Table \ref{tab:floor}.} If we assume  the optimistic noise floors of Table \ref{tab:floor}, we estimate that an atmosphere can be detected by JWST using the CO$_2$ absorption at $4.3\ \mu m$ for:

\begin{itemize}
    \item A modern Earth-like atmosphere with NIRSpec prism from  7 (TRAPPIST-1g) to 13 (TRAPPIST-1e \added{\footnote{Note that the relative transit depth of the CO$_2$ at $4.3\ \mu m$ for TRAPPIST-1e (47~ppm) is just below the $5\ \sigma$ noise floor value of 50~ppm but we assumed it detectable considering the uncertainty of this calculation.}}) transits at $3\ \sigma$ or 19 to 35 transits at $5\ \sigma$.
    \item An Archean Earth-like atmosphere with NIRSpec prism from  3 (TRAPPIST-1g) to 8 (TRAPPIST-1e) transits at $3\ \sigma$ or 9 (TRAPPIST-1g) to 23 (TRAPPIST-1e) transits at $5\ \sigma$.
    \item A CO$_2$ rich atmosphere (1 and 10~bars) with NIRSpec prism  from  5 (TRAPPIST-1g) to 10 (TRAPPIST-1e) transits at $3\ \sigma$ or  between 13 (TRAPPIST-1g)  to 28 (TRAPPIST-1e) transits at $5\ \sigma$.
\end{itemize}

Considering the conservative noise floors \citep{Greene2016} we estimate that an atmosphere can be detected (only at $3\ \sigma$)  using the CO$_2$ absorption at $4.3\ \mu m$ for:

\begin{itemize}
    \item A modern Earth-like atmosphere with NIRSpec prism from  7  transits for TRAPPIST-1f and TRAPPIST-1g, no detection for 1e.
    \item An Archean Earth-like atmosphere with NIRSpec prism for TRAPPIST-1g from  3 transits (TRAPPIST-1g) to 8 transits (TRAPPIST-1e).
    \item A CO$_2$ rich atmospheres (1 and 10~bars) with NIRSpec prism are detectable from 5 (TRAPPIST-1f at 1~bar) to 7 (TRAPPIST-1e at 1 bar) transits. CO$_2$ transit depth for TRAPPIST-1e at 10~bars is below the noise floor.
\end{itemize}

Note that in the MIRI range, the higher value of the noise floors and/or the  number of transits greater than 100 compromise the chance of detecting an atmosphere for the TRAPPIST-1 planets in the HZ with this instrument during JWST lifetime.    
Concerning  gases others than CO$_2$ such as O$_2$, O$_3$, CH$_4$ or even H$_2$O, according to our simulated atmospheric scenarios, none of them are detectable during JWST at a $5\ \sigma$ confidence level even for a photon-limited (white noise) estimation. 

\subsection{Water features}
\begin{figure}[h!]
\centering
\resizebox{9cm}{!}{\includegraphics{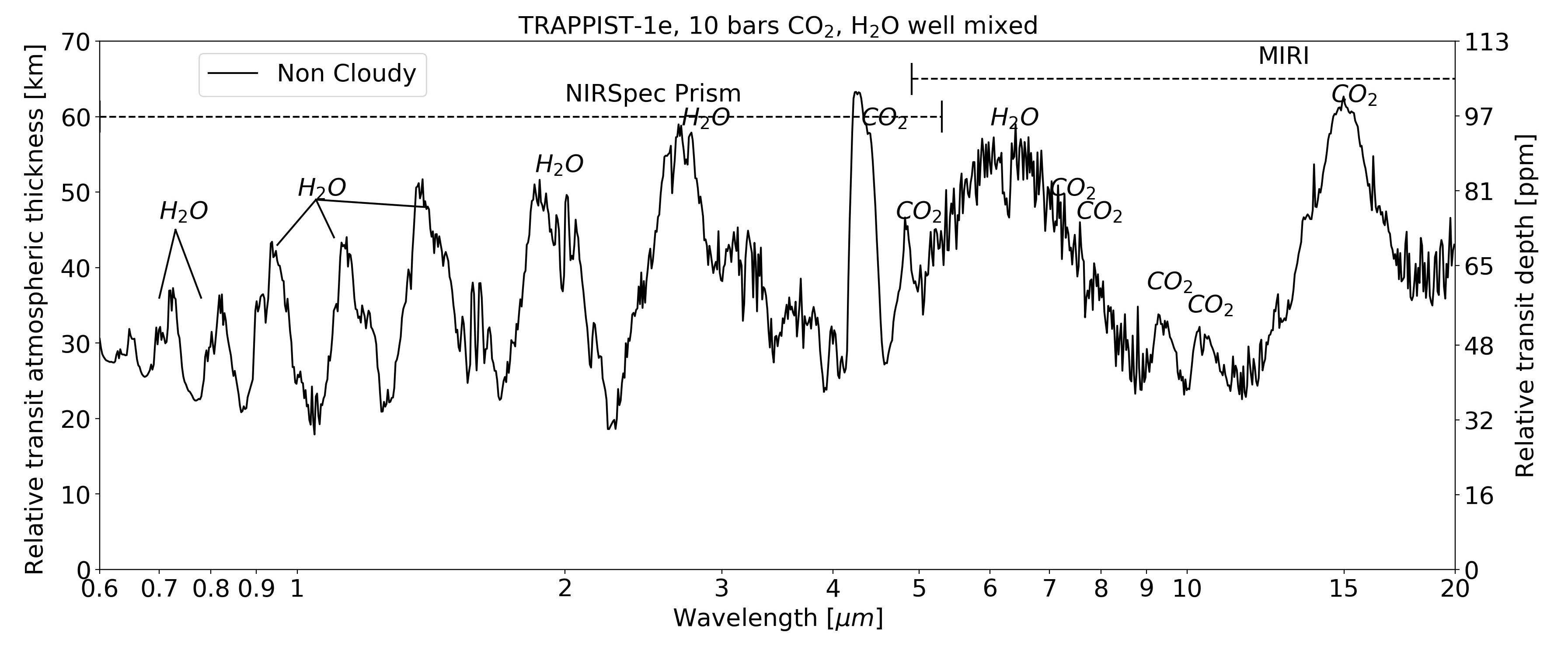}}
\caption{TRAPPIST-1e transmission spectrum for the 10 bars of CO$_2$ atmosphere for which H$_2$O has been forced to be vertically well-mixed. In transmission spectroscopy, H$_2$O are much stronger when H$_2$O is vertically well-mixed than when H$_2$O is confined near the surface.}
\label{fig:runaway}
\end{figure}

In this work, we have shown that water lines are challenging to detect from JWST transmission spectroscopy for habitable planets in the TRAPPIST-1 system (or equivalent system of planets in the HZ of an (ultra-cool) M dwarf). GCM simulations of those worlds show that, water vapor stays confined in the lower atmosphere of the planets, namely in the troposphere. Nevertheless, layers near the surface are  warmer, leading to an increasing infrared opacity of the water continuum and shallow lines. In this situation, even a small amount of well-mixed CO$_2$ in the atmosphere is enough to largely dominate over H$_2$O lines and hide them (like at $2.7\ \mu m$). As a result, none of the water vapor lines is larger than the presumed $3\ \sigma$ or $5\ \sigma$ noise floors (see Table \ref{tab:floor}). To have a large water mixing ratio through the whole atmospheric column would require either a moist greenhouse or runaway climate or a very low atmospheric pressure, in which case the atmospheric cold trap is suppressed in particular in the substellar region, and the H$_2$O mixing ratio can remain high in the upper atmosphere \citep{Turbet2016}. \\
To represent how the confinement of H$_2$O near the surface affects its detectability  we have considered the following thought experiment for  TRAPPIST-1e with 10 bars of CO$_2$, clear sky: The average atmospheric H$_2$O vapor mixing ratio ($47\ \%$), confined below 20~km is now well-mixed horizontally and vertically. \added{While this is maybe a rather unrealistic scenario, it helps to understand how H$_2$O mixing through the atmosphere impacts the strengh of the water lines.} The resulting JWST transmission spectra  is showed in Fig. \ref{fig:runaway}. The H$_2$O lines are now much stronger. In the NIRSpec range at a resolving power of 30 it is difficult to find a H$_2$O line not blended by CO$_2$ except for the shorter wavelengths. At $0.95\ \mu m$  the  H$_2$O feature line reach up to 32~ppm and 89 transits would be needed to achieve a $5\ \sigma$ detection but the transit depth is below the 50~ppm  noise floor at $5\ \sigma$. If one consider the significance of a $3\ \sigma$ detection this absorption line would be detectable in 32 transits.  Note that we do not include clouds here because we could not predict how cloud will form in such atmosphere with running a water-loss  simulations. However, they are expected to severely affect the detectability of the H$_2$O features. In general, this result demonstrates that the use of 3-D climate models (taking self-consistently into account the effect of clouds and sub-saturation) is crucial to correctly evaluate the detectability of condensible species such as water.\\

%FIGURE 18

Water vapor in the atmosphere intrinsically leads to water cloud formation either in the liquid or ice phase. Clouds are formed where the majority of the water vapor is in a non-runaway atmosphere, and they partially block the transmitted light, flattening the spectrum especially for H$_2$O. Well mixed species such as CO$_2$ are less impacted because enough of it remains above the cloud deck.\\

\begin{figure}[h!]
\centering
\resizebox{9cm}{!}{\includegraphics{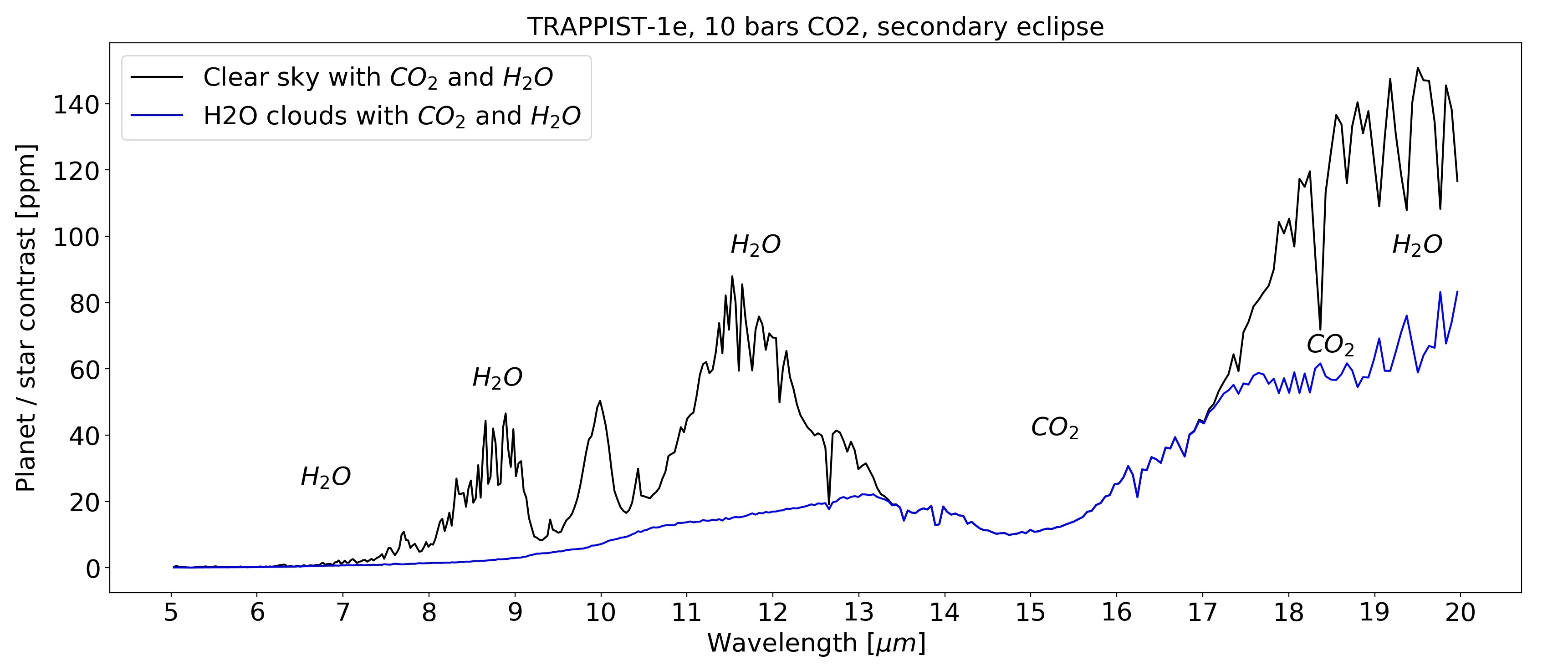}}
\caption{Emission spectrum from the secondary eclipse with MIRI (R=300) for TRAPPIST-1e with an atmosphere consisting of a surface pressure of 10 bars of CO$_2$.}
\label{fig:emission}
\end{figure}

Concerning emission spectroscopy, this technique is more sensitive to hot planets near the star \citep{Morley2017}. The hottest simulations we have performed are TRAPPIST-1e at 10 bars of CO$_2$. Figure \ref{fig:emission} shows the emission spectrum for MIRI from 5 to $20\ \mu m$ with a R=300 for the secondary eclipse. Below $5\ \mu m$, the contribution of the star removes the signal. The black curve shows the clear-sky spectrum and the blue curve shows the cloudy-sky spectrum.  While few strong H$_2$O lines are present here in the clear-sky case, once clouds are considered those lines are flattened. The H$_2$O lines that are less impacted are near $20\ \mu m$, with a thermal contrast of 25~ppm, but this is much smaller than the noise in the MIRI and therefore will be hard to detect. Beyond $20\ \mu m$ clouds become more transparent and the planet/star contrast increases and the noise increases dramatically as well. Therefore, it seems that emission 
spectroscopy with JWST is not helpful  to detect H$_2$O lines, in agreement with \cite{Lustig_Yaeger2019}.\\

Reflection spectroscopy may be a better option to probe water vapor lines because, contrary to the transmission spectroscopy for which the starlight is transmitted through the terminator of the planet, it could probe the disk of the planet where clouds could be absent in some regions, at various phases. Also  reflection spectroscopy can probe the lowest level of the atmosphere where most of the water resides. However, the small inner working angle (IWA) of the instrument on future direct imaging missions such as LUVOIR or HabEX would prevent the observation of such  compact system like TRAPPIST-1.

The combination of the two effects: 1) the water vapor confined in the low atmosphere and 2) the cloud opacities imply that the detection of water vapor lines may be challenging to detect for  planets in the habitable zone of TRAPPIST-1 or equivalent system. \\

%FIGURE 19

\section{Conclusions}\label{sec:conclusions}

In this work we have successfully connected a global circulation model (LMD-G) to a 1-D photochemical model (Atmos) and then applied a spectrum generator and noise model to estimate the detectability of gas species in a realistic set of possible atmospheres for the TRAPPIST-1 planets in the HZ. This has led to a consistent estimation of the cloud coverage along with the atmospheric temperatures and water profiles of tidally-locked planets around M dwarfs. However, the haze formation and photochemistry have been limited to the terminator region only, while a more consistent way would be to fully couple the GCM and the photochemical model. Also, no ocean heat transport has been considered (LMD-G does not have this feature yet) but that should not qualitatively impact our results. \added{This coupling would lead to more clouds migrating} toward the terminator with OHT enabled. These effects will be investigated in future studies.\\

We have seen that the Archean Earth-like atmospheres offer habitable conditions (ice-free surface) for TRAPPIST-1e and TRAPPIST-1f while only TRAPPIST-1e is habitable if an atmosphere  with boundary conditions based on the modern Earth  is considered. The CO$_2$ atmospheres lead to very high surface temperatures. TRAPPIST-1e is fully habitable at 1~bar of CO$_2$ while TRAPPIST-1f is an eye-ball planet (TRAPPIST-1g atmosphere collapses at 1~bar of CO$_2$). At 10~bars of CO$_2$, TRAPPIST-1e and 1f surface temperatures are so high that the oceans should evaporate leading to desiccated planets. On the other hand, TRAPPIST-1g  holds a few habitable ice-free spots near the substellar point.\\

Using the simulated JWST transmission spectroscopy, we found that an atmosphere with varying concentrations of CO$_2$ would be detectable for all habitable atmosphere configurations presented in this work in less than 15 transits  at  $3\ \sigma$ or less than 35 transits at  $5\ \sigma$ with NIRSpec. Nevertheless, CO$_2$ is expected to be an abundant gas in an exoplanet atmosphere owing to its  large abundance in the rocky planet atmospheres of our Solar system and to its high molecular weight making it more resistant to atmospheric escape. This number of transit observations is reasonably achievable during the lifetime of JWST.
Unfortunately, we did not find any gas other than CO$_2$ to be detectable during JWST nominal life time or in less than 100 transits. Overall, it appears that NIRSpec performed better in terms of signal-to-noise ratio and minimizes the number of transits in comparison to MIRI. However, if hazes are detected on these planets, MIRI (in its shortest wavelengths between 5 to 10 $\mu m$) may perform better because the haze opacity is much lower than within the NIRSpec range.
This study also suggests that it is very challenging to detect water lines for habitable planets orbiting ultra-cool dwarf stars such as TRAPPIST-1. Indeed, water mostly remains confined to the lower levels of the atmosphere with higher IR opacity leading to shallow lines (well below the noise floor of next space observatories), very often blended by stronger lines of the well-mixed CO$_2$. Water may be mixed through the entire atmospheric column if the planets are in a moist greenhouse state, but it will require either a very high amount of GHG (greater than 10~bars) and/or an instellation larger than the one received in the habitable zone and/or a very thin atmosphere suppressing cold traps. In addition, water vapor in the atmosphere implies the formation of water clouds blocking the transmitted light and leading to the flattening of water lines. \\

\added{Many effects are in competition to determine which of the three planets in the HZ offers the best chance of detection of an atmosphere. From TRAPPIST-1e to 1g, the planets get colder (decreasing the transit depth) but increase\deleted{s} in size (increasing the transit depth). TRAPPIST-1e has the largest gravity (decreasing the transit depth) following by 1g and 1f. Our simulations suggest that if a modern Earth-like atmosphere is present on the TRAPPIST-1 HZ planets, TRAPPIST-1e would be the most cloudy, while TRAPPIST-1f would be the most hazy for an Archean Earth-like atmosphere. Farther away, TRAPPIST-1g would have the lowest cloud and haze coverage. The atmospheric refraction also increases from TRAPPIST-1e to 1g, rising the continuum level to higher altitudes and therefore reducing the relative transit depths of the absorption lines. Finally, the frequency of observable transits during JWST's nominal lifetime decreases from TRAPPIST-1e (85), to 1f (55) and 1g (42). Overall, it appears that  TRAPPIST-1g offers the most favorable conditions for a detection of an atmosphere, using the CO$_2$ line at $4.3\ \mu m$.  We also found that larger ice-free surfaces lead to more clouds formed,  themselves significantly hiding habitability markers (such has water vapor absorption lines). It is then possible that habitability would be more feasibly  detectable if the planet is habitable locally \added{rather} than globally.}

\acknowledgments
T. Fauchez, G. Villanueva, G. Arney, R. Kopparapu, A. Mandell and S. Domagal-Goldman acknowledge support from GSFC Sellers Exoplanet Environments Collaboration (SEEC), which is funded in part by the NASA Planetary Science Division’s Internal Scientist Funding Model.\\
This project has received funding from the European Research Council (ERC) under 
the European Union’s Horizon 2020 research and innovation program (grant agreement No. 724427/FOUR ACES). This project has received funding from the European Union’s Horizon 2020 research and innovation program under the Marie Sklodowska-Curie Grant Agreement No. 832738/ESCAPE. \added{We would like to thank the anonymous reviewer for comments that greatly improved our manuscript. We also thank Amy Houghton from USRA for her  proofreading of the manuscript.}  \\



\software{Atmos \citep{Arney2016}, 
          LMD-G \citep{Wordsworth2011},
          PSG \citep{Villanueva2018}
          }



%\appendix

%\section{Appendix information}

 \bibliography{ref_latex.bib}
\listofchanges
%\section{Figures}

\end{document}